\documentclass[12pt,preprint2]{aastex}
\usepackage{amsmath}

\shorttitle{STELLAR PARAMETERS AND OXYGEN ABUNDANCES OF G--K DWARFS}
\shortauthors{Takeda et al.}

\begin{document}


\title{SPECTROSCOPIC DETERMINATION OF STELLAR PARAMETERS AND 
OXYGEN ABUNDANCES FOR HYADES/FIELD G--K DWARFS
}



\author{
{\sc Yoichi Takeda}\altaffilmark{1,2}
 {\sc and}
{\sc Satoshi Honda}\altaffilmark{3}
}
\altaffiltext{1}{National Astronomical Observatory, 2-21-1 Osawa, 
Mitaka, Tokyo 181-8588, Japan}
\email{takeda.yoichi@nao.ac.jp, honda@nhao.jp}
\altaffiltext{2}{SOKENDAI, The Graduate University for Advanced Studies, 
2-21-1 Osawa, Mitaka, Tokyo 181-8588, Japan}
\altaffiltext{3}{Nishi-Harima Astronomical Observatory, Center for Astronomy,\\
University of Hyogo, 407-2 Nishigaichi, Sayo-cho, Sayo, Hyogo 679-5313, Japan}


\begin{abstract}
It has been occasionally suggested that Fe abundances of K dwarfs derived from Fe~{\sc i} 
and Fe~{\sc ii} lines show considerable discrepancies and oxygen abundances determined 
from high-excitation O~{\sc i} 7771--5 triplet lines are appreciably overestimated
(the problem becoming more serious towards lower $T_{\rm eff}$), which however 
has not yet been widely confirmed. 
With an aim to clarify this issue, we spectroscopically determined the atmospheric 
parameters of 148 G--K dwarfs (Hyades cluster stars and field stars) by 
assuming the classical Fe~{\sc i}/Fe~{\sc ii} ionization equilibrium as usual, 
and determined their oxygen abundances by applying the non-LTE spectrum fitting 
analysis to O~{\sc i} 7771--5 lines.
It turned out that the resulting parameters did not show any significant 
inconsistency with those determined by other methods (for example, the mean 
differences in $T_{\rm eff}$ and $\log g$ from the well-determined solutions
of Hyades dwarfs are mostly $\lesssim 100$~K and $\lesssim 0.1$~dex).
Likewise, the oxygen abundances of Hyades stars are around [O/H]~$\sim +0.2$~dex
(consistent with the metallicity of this cluster) without exhibiting any 
systematic $T_{\rm eff}$-dependence. 
Accordingly, we conclude that parameters can be spectroscopically evaluated 
to a sufficient precision in the conventional manner (based on the 
Saha--Boltzmann equation for Fe~{\sc i}/Fe~{\sc ii}) and oxygen abundances 
can be reliably determined from the O~{\sc i} 7771--5 triplet for K dwarfs  
as far as stars of $T_{\rm eff} \gtrsim 4500$~K are concerned. 
We suspect that previously reported strongly $T_{\rm eff}$-dependent discrepancies 
may have stemmed mainly from overestimation of weak-line strengths and/or improper 
$T_{\rm eff}$ scale. 
\end{abstract}


\keywords{stars: abundances --- stars: atmospheres --- stars: fundamental parameters 
--- stars: late-type --- open clusters and associations: individual (Hyades)}



\section{INTRODUCTION}

It has been reported by several investigators that significant difficulties are 
involved in the spectroscopic analysis of lower main-sequence stars of late G 
to K-type (hereinafter we refer to this star group simply as ``K dwarfs''). 
That is, the Fe abundances derived from lines of neutral and ionized 
stages (Fe~{\sc i} and Fe~{\sc ii}) are not consistent with each other (generally 
the latter is larger than the former), and this Fe~{\sc ii} vs. Fe~{\sc i} discrepancy 
becomes progressively more serious as the effective temperature ($T_{\rm eff}$) 
is lowered. See, e.g., Allende Prieto et al. (2004, cf. their Fig.~8); Kotoneva 
et al. (2006, cf. their Fig.~9), and Luck (2017, cf. his Fig.~1) for field stars; 
King \& Schuler (2005; cf. their Fig.~4) for UMa moving group stars; Yong et al. 
(2004, cf. their Fig.~4) and Schuler et al. (2006b, cf. their Fig.~3) for Hyades 
cluster stars; Schuler et al. (2010; cf. their Fig.~1) for Pleiades cluster stars.
Whichever reason is relevant for this trend (e.g., substantial non-LTE 
overionization effect related to stellar activity; cf. Takeda 2008), 
it must have a large impact if it is real, given the paramount importance of Fe lines 
in stellar spectroscopy.
For example, the widely used method of determining the atmospheric parameters of
solar-type stars based on Fe~{\sc i} and Fe~{\sc ii} lines (which makes use of
the excitation equilibrium of Fe~{\sc i} and ionization equilibrium of 
Fe~{\sc i}/Fe~{\sc ii}; e.g., Takeda et al. 2002) would hardly be applicable to K dwarfs, 
since classical 1D plane-parallel model atmospheres would be no more valid for them.

However, some doubt still remains regarding whether this effect is really so important.
Wang et al. (2009) carried out spectroscopic analysis of 30 nearby lower main-sequence
stars at 4700~$\lesssim T_{\rm eff} \lesssim$~5400~K. They could not confirm the 
appreciable $T_{\rm eff}$-dependent systematic discrepancy reported by Kotoneva et al. 
(2006), but found a reasonable consistency between Fe~{\sc i} and Fe~{\sc ii} 
abundances to a level of $\lesssim 0.1$~dex  (cf. Fig.~5 therein).
Furthermore, Aleo et al. (2017) conducted an extensive examination on this
alleged ``Fe abundance anomaly in K dwarfs'' by carefully determining the Fe 
abundances from lines of neutral and ionized stages for 63 wide binary stars
and 33 Hyades stars at 4300~$\lesssim T_{\rm eff} \lesssim$~6100~K.
Their important finding is the importance of line-blending effect for certain 
Fe~{\sc ii} lines, which becomes prominent for K dwarfs of lower $T_{\rm eff}$
where Fe~{\sc ii} lines are weaker while lines of neutral metals get stronger. 
By removing these lines, they found that the Fe~{\sc ii}--Fe~{\sc i} 
discrepancy is appreciably mitigated; e.g., for Hyades stars, only $\sim 0.1$~dex 
at 4500~K~$\lesssim T_{\rm eff}$, though increasing to $\sim 0.3$~dex at further lower 
$T_{\rm eff}$ of $\sim 4300$~K (cf. Fig.~9 therein). 
Likewise, Tsantaki et al. (2019) very recently performed a detailed study
on the Fe ionization equilibrium based on the spectra of 451 FGK-type stars
(subsample of HARPS GTO planet survey program) and also arrived at the conclusion
that unresolved line blending is probably the main reason for the apparent overabundance
of Fe~{\sc ii}. They showed that $T_{\rm eff}$-independent consistent results
could be obtained by rejecting suspicious Fe~{\sc ii} lines.  
These two recent investigations suggest that the considerably 
large  Fe~{\sc ii}--Fe~{\sc i} disagreement reported in previous studies (e.g., as much 
as $\sim$~0.5--0.6~dex at $T_{\rm eff} \sim 4500$~K for the case of Hyades K dwarfs; cf.
Yong et al. 2004, Schuler et al. 2006b) is likely to be due to their inadequate choice
of blending-affected Fe~{\sc ii} lines, leading to a significant overestimation of 
Fe~{\sc ii} abundances. 

This revelation reminded us of a similar problem related to oxygen 
abundance determination for K dwarfs. That is, the widely used high-excitation 
O~{\sc i} 7771--5 triplet lines tend to result in erroneously overestimated 
abundances (being progressively more serious with a decrease in $T_{\rm eff}$), 
which was reported in several studies on open cluster stars:
UMa moving group (King \& Schuler 2005; cf. their Fig.~4), M~34 as well as 
Pleiades (Schuler et al. 2004; cf. their Fig.~1 and Fig.~2), Hyades (Schuler et al. 
2006a; cf. their Fig.~3), and NGC~752 (Maderak et al. 2013; cf. their Fig.~5).  
Actually, this effect of abundance anomaly they found was surprisingly large, 
because [O/H] values (oxygen abundance relative to the Sun) of K dwarfs 
derived from O~{\sc i} 7771--5 lines turned out to be unreasonably higher than
those of G dwarfs by as much as $\lesssim 1$~dex, despite that they should have 
similar values for stars belonging to the same cluster. 

Although their investigations were based on the assumption of LTE, the non-LTE effect 
evaluated in the standard manner using classical model atmospheres (see, e.g., 
Takeda 2003) can not explain this apparently large overabundance of [O/H], 
because non-LTE correction is strength-dependent and almost negligible for K dwarfs, 
where high-excitation O~{\sc i} 7771--5 lines are considerably weak because 
of lower $T_{\rm eff}$. So, if this is real, it might be due to some kind of 
non-classical activity-related phenomenon such as the intensification caused by 
chromospheric temperature rise (cf. Takeda 2008).
However, in view of the similarity to the case of Fe abundance discrepancy 
(in the sense that considerably weak Fe~{\sc ii} and O~{\sc i} lines 
are involved for the anomalous abundances seen in K dwarfs), this problem on the 
reliability of O~{\sc i} 7771--5 triplet may be worth reinvestigation.

This situation motivated us to revisit these ``spectroscopic K dwarf problems on
Fe and O abundances'' based on the spectral data for a large sample of G--K dwarfs 
(47 Hyades stars and 101 field stars). 
Our approach is simply to apply the standard method of analysis adopted in our previous
studies to all these sample stars and see if any unreasonable result (such as 
suggesting the breakdown of classical modeling) comes out or not.
More precisely, what we want to do and clarify in this investigation is as follows: 
\begin{itemize}
\item
We determine the atmospheric parameters of each program star in the conventional
manner from the equivalent widths of Fe~{\sc i} and Fe~{\sc ii} lines while 
assuming the LTE (Saha--Boltzmann equation) as done by Takeda et al. (2005). 
If classical treatment is not valid for K dwarfs, the resulting parameters 
would reveal some kind of $T_{\rm eff}$-dependent inconsistency between G and K dwarfs.  
In this context, Hyades stars should play an especially important role, because
their parameters are known to a sufficient precision. Comparison of our spectroscopically
determined parameters and those already established by other methods would make a decisive
touchstone.\footnote{Although Takeda (2008) once conducted this reliability test of 
spectroscopic parameters using Hyades stars of 5100~K~$\lesssim T_{\rm eff} \lesssim$~6200~K,
few K-type dwarfs of our main interest were included unfortunately, Besides, the spectra 
used at that time (later analyzed also by Takeda et al. 2013) were limited to the wavelength 
range of $\sim$~6000--7200~\AA\, and thus not necessarily sufficient in view of 
the number of available Fe~{\sc i} and Fe~{\sc ii} lines. Therefore, we decided to 
redo this task by using new observational data of wider wavelength coverage.} 
\item
By using the model atmospheres corresponding to such determined atmospheric parameters,
we then evaluate the oxygen abundance for each star by applying the spectrum-fitting 
method to O~{\sc i} 7771--5 (see, e.g., Takeda et al. 2015), where the non-LTE
effect was taken into account according to Takeda's (2003) calculation. 
Again, Hyades stars can serve as a good testbed, because they are considered to
have practically the same O abundances. Are the resulting [O/H] values consistent 
with each other? Or do they show such considerable $T_{\rm eff}$-dependent disagreement
as concluded by Schuler et al. (2006a)?  This must be an interesting check.
In addition, it is worthwhile to examine the behavior of [O/Fe] ($\equiv$[O/H]$-$[Fe/H]) 
ratios with a change in [Fe/H] (metallicity) obtained for field stars. Is the [O/Fe] vs. [Fe/H] 
diagram obtained from O~{\sc i} 7771--5 lines for K dwarfs consistent with that
derived by Takeda \& Honda (2005) for F--G stars with the same triplet?  
This can be another touchstone for judging the reliability of these high-excitation 
O~{\sc i} lines in context of oxygen abundance determination for K dwarfs. 
\end{itemize}

\section{OBSERVATIONAL DATA}


As the target stars of this investigation, we adopted a sample of 148 dwarfs 
(consisting of 47 Hyades cluster stars and 101 field stars), which are in the 
apparent magnitude range of $V \sim$~5--10.
Regarding Hyades stars, since we intended to cover a rather wider range of spectral type
(in order to clarify the $T_{\rm eff}$-dependence), main-sequence stars in the color 
range of $0.5 \lesssim B-V \lesssim 1.2$ (corresponding to late F through mid K) were 
selected from de Bruijne et al.'s (2001) list.
As to field stars, we mainly invoked Kotoneva et al.'s (2002) list of G--K dwarfs,
from which 99 stars in the color range of $0.7 \lesssim B-V \lesssim 1.2$ 
(corresponding to mid--late G through mid K) were chosen. In addition, in order to
reinforce the sample of mid-K stars, 61~Cyg~A and $\xi$~Boo~B (both having 
$B-V \sim 1.2$) were also included.
The basic data of these 148 stars are summarized in Table~1 (and in ``tableE1.dat'' 
of the online material). The $M_{V}$ vs. $B-V$ diagram for the program stars
is shown in Figure~1a.


Our spectroscopic observations for 118 stars were done in 4 runs of 2010--2011 
(2010 April/May, 2010 August, 2010 November/December, and  2011 November) by 
using the HIDES (HIgh Dispersion Echelle Spectrograph) placed at the coud\'{e} 
focus of the 188 cm reflector at Okayama Astrophysical Observatory.
Equipped with three mosaicked 4K$\times$2K CCD detectors
at the camera focus, HIDES enabled us to obtain an echellogram covering 
$\sim$~5100--8800~$\rm\AA$ with a resolving power of $R \sim 67000$. 
The observations for the remaining 30 stars were done on 2014 September 9 
with the HDS (High Dispersion Spectrograph) placed at the Nasmyth platform 
of the 8.2-m Subaru Telescope, by which high-dispersion spectra 
with a resolution of $R \simeq 80000$ covering 
$\sim$~5100--7800~$\rm\AA$ (with two 4K$\times$2K CCDs) were obtained.
The observed dates for each of the program stars are given in ``tableE1.dat''.


The reduction of the spectra (bias subtraction, flat-fielding, 
scattered-light subtraction, spectrum extraction, wavelength 
calibration, and continuum normalization) was performed by using 
the ``echelle'' package of the software IRAF\footnote{
IRAF is distributed by the National Optical Astronomy Observatories,
which is operated by the Association of Universities for Research
in Astronomy, Inc. under cooperative agreement with the National 
Science Foundation.} in a standard manner. 
If a few consecutive exposures were done for a star in a night, 
we co-added these to improve the signal-to-noise ratio.
The average S/N of the finally resulting spectrum is typically 
$\sim$~100--300 for most cases. 

\section{STELLAR PARAMETERS}

\subsection{Atmospheric Parameters Based on Fe Lines}

The four parameters [$T_{\rm eff}$ (effective temperature), $\log g$ 
(logarithmic surface gravity), $v_{\rm t}$ (microturbulent velocity dispersion), 
and [Fe/H] (Fe abundance relative to the Sun)] were spectroscopically 
determined from the equivalent widths ($W_{\lambda}$) of Fe~{\sc i} 
and Fe~{\sc ii} lines based on the principle and algorithm described 
in Takeda et al. (2002), which requires that
(i) Fe~{\sc i} abundances do not depend upon $\chi_{\rm low}$
(lower excitation potential), (ii) mean Fe~{\sc i} and Fe~{\sc ii} 
abundances are equal, and (iii) Fe~{\sc i} abundances do not depend upon 
$W_{\lambda}$, while assuming that LTE Saha--Boltzmann equation holds.

The measurement of $W_{\lambda}$ for each Fe line (selected from the line list
of Takeda et al. 2005) was done by the Gaussian-fitting method in most cases 
(though special function constructed by convolving the rotational 
broadening function with the Gaussian function was used for several cases 
of appreciably large rotational velocity). 
In order to avoid measuring inadequate lines affected by blending, 
we carried out measurements on the computer display, while comparing 
the stellar spectrum with Kurucz et al's (1984) solar spectrum 
and examining the theoretical strengths of neighborhood lines computed 
with the help of Kurucz \& Bell's (1995) atomic line data.

Practically, we applied the program TGVIT (Takeda et al. 2005; cf. Sect. 2 therein), 
to the measured $W_{\lambda}$'s of Fe~{\sc i} and Fe~{\sc ii} lines. 
As done in Takeda et al. (2005), we restricted to using lines satisfying 
$W_{\lambda} \le 100$~m$\rm\AA$ and those showing abundance deviations from the mean 
larger than $2.5 \sigma$ were rejected. The typical numbers of finally adopted lines 
are 72--235 (mean = 208) for Fe~{\sc i} and 5--22 (mean = 14) for Fe~{\sc ii} 
(number of available lines is smaller for stars showing broader lines).
The resulting final parameters are presented in Table~1 and ``tableE1.dat''. 
The internal statistical errors (cf. Section~3.2 of Takeda et al. 2002) involved 
with these solutions are in the range of
10--85~K (mean = 24~K) for $T_{\rm eff}$, 
0.02--0.26~dex (mean = 0.06~dex) for $\log g$, 
0.1--0.5~km~s$^{-1}$ (mean = 0.2~km~s$^{-1}$) for $v_{\rm t}$,
and 0.01--0.07~dex (mean = 0.03~dex) for [Fe/H].
The detailed data of $W_{\lambda}$ and $A$(Fe) (Fe abundances corresponding 
to the final solutions) for each star are given in ``tableE2.dat'' of the 
online material.

\subsection{Trends and Mutual Correlations}

These spectroscopically determined $T_{\rm eff}$ values are plotted against 
$B-V$ and $M_{V}$ in Figures~1b and 1c, where we can see that they are well 
correlated with each other. The color-dependence of [Fe/H] depicted
in Figure~1d indicates the near-constancy of [Fe/H] for Hyades stars 
and a tendency of decreasing [Fe/H] towards a bluer $B-V$ for field stars 
(consistent with  Fig.~8 of Kotoneva et al. 2002).
Figure~1e shows the comparison of our spectroscopic [Fe/H] with the photometric
metallicity ([Fe/H]$^{\rm photo}$) determined by Kotoneva et al. (2002) based on 
the position in the color--magnitude diagram, which shows a reasonable correlation
between these two (though their [Fe/H]$^{\rm photo}$ tends to be somewhat lower).  

In Figures 2a--2c are plotted $\log g$, $v_{\rm t}$, and [Fe/H] against $T_{\rm eff}$,
where the results of 160 dwarfs/subgiants (of mostly F--G type) determined by 
Takeda et al. (2005) are also shown for comparison.
We can see from Figure~2a (where theoretical $\log g$ vs. $\log T_{\rm eff}$ relations 
are also depicted) that most of our program stars occupy consistent positions 
as main-sequence stars. However,  deviations (i.e., underestimation of $\log g$) 
begins to appear towards low $T_{\rm eff}$ end, which means that precision of 
$\log g$ tends to gradually deteriorate as $T_{\rm eff}$ is lowered
below $\lesssim 5000$~K. (see Section~5.1). 

Regarding microturbulence, meaningless negative $v_{\rm t}$ values were 
obtained for two considerably metal-poor stars HIP~057939 ($-0.10$~km~s$^{-1}$)
and HIP~098792 ($-0.18$~km~s$^{-1}$), which is due to the result of extrapolation.
In actual determination of oxygen abundance (cf. Section~4),  
we tentatively assigned $v_{\rm t} = 0.5$~km~s$^{-1}$ for these stars.
We also note that two stars (HIP~093926, HIP~092919) show anomalously high
$v_{\rm t}$ values ($\sim 2$~km~s$^{-1}$), which must be related to the fact
that these stars show exceptionally broad lines indicative of higher rotation.
It is interesting to note in Figure~2b that, while the $v_{\rm t}$ results determined 
for 101 field stars (blue circles) tend to decrease as $T_{\rm eff}$ is lowered 
as a natural continuation of the trend derived by Takeda et al. (2005) (represented 
by green dots), those obtained for 47 Hyades stars (red crosses) appear to be almost
independent upon $T_{\rm eff}$ and nearly flat at $\sim 1$~km~s$^{-1}$. This may 
suggest a possibility that $v_{\rm t}$ could be somehow influenced by stellar age 
or activity, because Hyades stars are comparatively younger and of higher activity.  

Figure~2c shows that the metallicities of Hyades stars are nearly constant
at [Fe/H]~$\sim 0.2$; i.e., the mean ($\pm$ standard deviation) is
$\langle$[Fe/H]$\rangle$ = 0.19 ($\pm 0.07$). This is slightly higher than 
the value of $\langle$[Fe/H]$\rangle$ = 0.11 ($\pm 0.08$) derived for F--G dwarfs
by Takeda et al. (2013), but consistent within permissible limits
in view of the fact that the published values of Hyades metallicity range 
at $0.1 \lesssim $~[Fe/H]~$\lesssim 0.2$.\footnote{
Takeda (2008) summarized the Hyades [Fe/H] values determined by 13 spectroscopic 
studies in 1971--2005 (cf. Fig.~32.8a therein), which are between +0.1 and +0.2
(the mean is +0.14 with a standard deviation of 0.03). The same argument almost holds 
for the more recent literature values, as summarized in Sect.~5.5 of
Dutra-Ferreira et al. (2016), who themselves derived two values of $+0.18\pm 0.03$ 
(method using well-constrained parameters) and $+0.14 \pm 0.03$ (classical method) 
for the average [Fe/H] value of dwarfs+giants in the Hyades cluster.}

Meanwhile, those for field G--K stars range mostly from $-0.7$ to $+0.3$ 
(like the case of 160 sample stars studied by Takeda et al. 2005), though
only HIP~057939 is distinctly metal-deficient ([Fe/H] = $-1.27$) compared to the others.
In connection with metallicity, it may be worth examining the population of 
our program stars. For this purpose, their kinematic parameters were computed by following 
the same procedure as adopted in Takeda (2007; cf. Sect.~2.2 therein), where the necessary 
data (equatorial coordinates, parallax, proper motions, and radial velocity\footnote{
We found that {\it Gaia} DR2 heliocentric radial velocities are consistent with those 
measured from our spectra for most of our program stars. The exceptional ones (showing 
differences more than 3~km~s$^{-1}$) are HIP~093926 ($-37.9$), HIP~013891 (+13.5), 
HIP~040419 ($-7.9$), HIP~104214 (+6.3), HIP~012158 ($-5.7$), and HIP~092919 ($-4.7$), 
where the parenthesized values are 
$V_{\rm rad}^{\rm hel}$({\it Gaia})$-$$V_{\rm rad}^{\rm hel}$(ours) (in km~s$^{-1}$).
These stars are likely to be spectroscopic binaries.}) were 
taken from those of {\it Gaia} DR2 (Gaia Collaboration et al. 2016, 2018) published as 
CDS/ADC Collection of Electronic Catalogues (No. 1345, 0, 2018) and available via SIMBAD.
The resulting orbital parameters and space velocity components relative to the 
Local Standard of Rest (LSR) are given in tableE1.dat of the online material.
The $z_{\rm max}$ (maximum separation from the galactic plane) vs. $V_{\rm LSR}$ 
(rotation velocity component) diagram usable for discriminating stellar population
is displayed in Figure~3a, which indicates that most of our target stars belong 
to the thin disk population (with a few exceptions such as HIP~057939 and 
HIP~082588 which may be of thick-disk population).
Figure~3b illustrates the correlation between the space velocity 
$|v_{\rm LSR}|$ ($\equiv \sqrt{U_{\rm LSR}^{2}+ V_{\rm LSR}^{2}+W_{\rm LSR}^{2}}$), 
and metallicity ([Fe/H]). Though the scatter is rather large, we can recognize 
that $|v_{\rm LSR}|$ tends to increase with a decrease in [Fe/H] as expected
It can also be seen that those two stars of apparent thick-disk population 
mentioned above show distinctly larger $|v_{\rm LSR}|$ (especially HIP~057939).

\section{OXYGEN ABUNDANCE DETERMINATION}

\subsection{Spectrum-Fitting Analysis}

We determine the oxygen abundances of 148 target stars from the O~{\sc i} 7771--5 
triplet feature as done in previous studies (e.g., Takeda et al. 2015).
Based on the atmospheric parameters determined in Section~3.1, the model atmosphere 
for each star was constructed by interpolating Kurucz's (1993) ATLAS9 model grid.  
We similarly evaluated the non-LTE departure coefficients for O corresponding to each 
model by interpolating the grid computed by Takeda (2003).

Abundance determination was carried out by using the spectrum-fitting technique
as done in Takeda et al. (2015), which establishes the most optimum solutions
accomplishing the best match between theoretical and observed spectra
by using the numerical algorithm described in Takeda (1995), 
while simultaneously varying the abundances of 
relevant key elements ($A_{1}$, $A_{2}$, $\ldots$), 
the macrobroadening parameter ($v_{\rm M}$),\footnote{
This $v_{\rm M}$ is the $e$-folding half-width of 
the Gaussian broadening function ($\propto \exp[-(v/v_{\rm M})^{2}]$),
which represents the combined broadening width of instrumental profile, 
macroturbulence, and rotational velocity.} 
and the radial-velocity (wavelength) shift ($\Delta \lambda$).

We selected 7770--7782~\AA\ as the wavelength region for fitting, 
which includes O~{\sc i} 7771--5 triplet lines and Fe~{\sc i} 7780 line 
as the conspicuous lines. Regarding the atomic data of spectral lines,  
the same values as used in Takeda et al. (2015) were used unchanged
for 3 lines of O~{\sc i} 7771--5 triplet and 6 lines of CN molecules
(cf. Table~2 therein). Otherwise, we invoked the data compiled
in the VALD database (Ryabchikova et al. 2015) for all lines 
included in this region (for example. $\log gf = +0.03$ was adopted for
the strong Fe~{\sc i} 7780.556 line of $\chi_{\rm low}$ = 4.47~eV).
We varied only $A$(O) and $A$(Fe) for the abundances to be adjusted,
while other elemental abundances (necessary for computing the background 
spectrum in this region) were fixed at the metallicity-scaled
values.\footnote{Although the abundances of CN and Nd were also varied
(in addition to O and Fe) in Takeda et al. (2015), we decided to fix them
in this study, because these line features are less significant for dwarfs 
compared to the case of giants. Note also that, since the role of 
$A$(Fe) is a fudge parameter to accomplish the satisfactory fit for the 
whole region, its solution was not used for deriving [Fe/H] of a star, for 
which we adopted the value determined from many Fe lines (cf. Section~3.1).} 
The non-LTE effect was taken into account for the O~{\sc i} 7771--5 lines.
Since the OAO/HIDES spectrum often suffers defects due to bad columns
of CCD in this region, we had to mask them occasionally.
The convergence of the solutions turned out fairly successful  
for all cases. How the theoretical spectrum for the converged 
solutions fits well with the observed spectrum for each star is 
displayed in Figure~4 (Hyades stars) and Figure~5 (field stars).

\subsection{Abundance-Related Quantities}

Next, with the help of Kurucz's (1993) WIDTH9 program 
(which had been considerably modified in various respects; e.g., inclusion 
of non-LTE effects, etc.), we computed the equivalent widths ($W_{7772}$,
$W_{7774}$, and $W_{7775}$) of three O~{\sc i} triplet lines (at 7771.944, 
7774.166, and 7775.388~\AA) inversely from the non-LTE abundance solution 
$A^{\rm N}$(O) (resulting from fitting analysis) along with the adopted 
atmospheric model and parameters.
Based on these $W$ values, the non-LTE ($A^{\rm N}$: essentially the same as
the fitting solution) and LTE ($A^{\rm L}$) oxygen abundances were then derived, 
from which the corresponding non-LTE corrections could be obtained as 
$\Delta \equiv A^{\rm N} - A^{\rm L}$.
In Table~1 (and also in ``tableE1.dat'') are presented [O/H] 
$(\equiv A^{\rm N} - 8.861)$,\footnote{Regarding the solar oxygen abundance, 
Takeda et al. (2015) derived $A^{\rm N}_{\odot} = 8.861$ (in the usual 
normalization of $A$(H)=12.00) as the non-LTE solar oxygen abundance 
with $\Delta_{7774} = -0.102$ and $W_{7774} = 63.2$~m\AA.
This solar $A^{\rm N}_{\odot}$ is the value obtained in the same manner
as adopted in this analysis (e.g., same line parameters, etc.),
which is necessary to accomplish the purely differential analysis for [O/H].
Although its absolute value is apparently larger than the recent solar oxygen 
abundance of 8.69 (Asplund 2009) and rather near to the old one
(e.g., 8.83 by Grevesse \& Sauval 1998), this difference does not matter here.
}
$W_{7774}$, $\Delta_{7774}$ (for the middle line of the triplet)

In order to estimate abundance errors caused by uncertainties
in atmospheric parameters, we derived six kinds of abundance variations
($\delta_{T+}$, $\delta_{T-}$, $\delta_{g+}$, $\delta_{g-}$, 
$\delta_{v+}$, and $\delta_{v-}$) for $A^{\rm N}$ by repeating the 
analysis on the $W_{7774}$ values while 
perturbing the standard atmospheric parameters interchangeably by 
$\pm 100$~K in $T_{\rm eff}$, $\pm 0.1$~dex in $\log g$, 
and $\pm 0.5$~km~s$^{-1}$ in $v_{\rm t}$ (which are larger than the internal 
statistical errors described in Section~3.1 but tentatively chosen
by considering possible systematic errors; cf. Section~5.1).

Errors ($\delta W$) due to random noises in the equivalent widths ($W$) 
were also estimated by invoking the relation derived by Cayrel (1988)
corresponding to the S/N ratio ($\sim$~100--200) measured for each star's 
spectrum in the neighborhood of O~{\sc i} triplet.
We then evaluated the abundances for each of the perturbed $W_{+} (\equiv W + \delta W)$ 
and $W_{-} (\equiv W - \delta W)$, from which the differences from the standard 
abundance ($A$) were derived as $\delta_{W+} (>0)$ and $\delta_{W-} (<0)$.

These $W_{7774}$, $\Delta_{7774}$, $A^{\rm N}$(O), $\delta_{W\pm}$, $\delta_{T\pm}$, 
$\delta_{g\pm}$, and $\delta_{v\pm}$ are plotted
against $T_{\rm eff}$ in panels (a)--(g) of Figure~6, respectively,
from which the following trends can be read. 
\begin{itemize}
\item
It can be seen that $W_{7774}$ progressively decreases as $T_{\rm eff}$ is lowered,
reflecting that the occupation number in the highly-excited lower level 
($\chi_{\rm low} = 9.15$~eV) of this transition is quite $T_{\rm eff}$-sensitive 
($\propto 10^{-5040 \chi_{\rm low}/T_{\rm eff}}$).
\item
Likewise, $|\Delta_{7774}|$ (absolute value of negative non-LTE correction)
declines with decreasing $T_{\rm eff}$, because of the close connection between 
$\Delta$ and $W$ (cf. Takeda 2003) for the O~{\sc i} 7771--5 triplet.
Accordingly, while non-LTE correction is still appreciable for late F--early G 
dwarfs ($\sim$~0.1--0.2~dex), it becomes practically negligible for K dwarfs of 
$T_{\rm eff} \lesssim 5000$~K. 
\item
The oxygen abundances ($A^{\rm N}$) do not show any clear $T_{\rm eff}$-dependence
for both Hyades and field stars. While the former are nearly constant on average
(though the scatter grows at $T_{\rm eff} \lesssim 5000$~K), the latter are 
diversified mostly in the range of $\sim$~8.5--9.2. 
\item
The mean of [O/H] for 47 Hyades stars is $\langle$[O/H]$\rangle$ = 0.11 ($\pm 0.09$),
Although this is slightly lower than the value ($\langle$[O/H]$\rangle$ = 0.22 $\pm 0.14$) 
derived by Takeda et al. (2013) for Hyades F--G dwarfs from O~{\sc i} 6156--8 lines,
we consider that both are reasonably consistent within the allowable range
(see Sect.~1 of Takeda et al. 2013 for a summary of published [O/H] values
in various literature).  
\item
Among the various sources of abundance errors, most important is $\delta_{T\pm}$   
(ranging from $\sim$~0.1~dex to $\sim 0.2$~dex or even more; especially large around 
lowest $T_{\rm eff}$) reflecting the high-excitation nature of O~{\sc i} triplet lines, 
while $\delta_{W\pm}$, $\delta_{g\pm}$ and $\delta_{v\pm}$ are comparatively 
insignificant (only several hundredths dex).
\end{itemize}

\section{DISCUSSION}

\subsection{Reliability of Spectroscopic Parameters}

As to whether stellar parameters of K dwarfs can be reliably determined based on Fe~{\sc i} 
and Fe~{\sc ii} lines, which was the first aim of this study (under the suspicion that
LTE ionization equilibrium of Fe~{\sc i}/Fe~{\sc ii} may break down), we can examine 
this problem by comparing the $T_{\rm eff}$ and $\log g$ values of Hyades dwarfs 
spectroscopically derived in Section~3.1 with those of de Bruijne et al. (2001), who 
made use of the theoretical color--magnitude relations along with the well-established 
luminosities from Hipparcos parallaxes. These comparisons are illustrated in Figure~7.

Figure~7a suggests that a rather satisfactory agreement is observed for $T_{\rm eff}$, though 
$T_{\rm eff}$(This study) tends to be is slightly higher than $T_{\rm eff}$(de Bruijne) 
by $\lesssim 100$~K (Figure~7c). The average 
$\langle \Delta T_{\rm eff}$(de Bruijne$-$This study)$\rangle$ is $-67 (\pm 50)$~K. 
Regarding $\log g$, we can see a tendency of $\log g$(This study) being smaller than 
$\log g$(de Bruijne) (Figure~7b). However, excepting two stars
(HIP~20762 and HIP~25639) the difference is within $\lesssim$0.1--0.2~dex (Figure~7d).
The average $\langle \Delta \log g$(de Bruijne$-$This study)$\rangle$ is $+0.06 (\pm 0.09)$~dex
(for all stars) or $+0.05 (\pm 0.06)$~dex (excluding two outliers). 
These $\Delta T_{\rm eff}$ and $\Delta \log g$ show a weak correlation (Figure~7e)
which is presumably because higher $T_{\rm eff}$ (enhancing ionization) is compensated 
by higher $\log g$ (suppressing ionization).

Considering the results of this test using Hyades G--K stars, we may conclude that our 
spectroscopically determined $T_{\rm eff}$ and $\log g$ do not suffer significant errors, 
which are determinable based on Fe~{\sc i} and Fe~{\sc ii} lines to typical precisions of 
$\lesssim 100$~K and $\lesssim 0.1$~dex under the assumption of LTE Saha--Boltzmann equation.
Admittedly, the tendency of slightly higher $T_{\rm eff}$ and lower $\log g$ in our
spectroscopic parameters may indicate a possibility of marginal overionization.
However, since $\Delta T_{\rm eff}$ as well as $\Delta \log g$ do not show any 
conspicuous dependence upon $T_{\rm eff}$, we can rule out the possibility of significant
$T_{\rm eff}$-dependent Fe~{\sc i}--Fe~{\sc ii} discrepancy progressively increasing 
towards lower $T_{\rm eff}$.
In this regard, our result is in favor of Aleo et al.'s (2017)
conclusion that such previously alleged considerable discordance between Fe~{\sc i} and 
Fe~{\sc ii} abundances in K dwarfs is largely due to improper inclusion of blended Fe~{\sc ii} 
lines and practically insignificant ($\lesssim 0.1$~dex) as long as stars of 
$T_{\rm eff} \gtrsim 4500$~K are concerned. 
We should note, however, that lowering of the precision is more or less unavoidable at
the low $T_{\rm eff}$ regime (see Section~3.2 in connection with the trend of $\log g$ vs. 
$T_{\rm eff}$ in Figure~2a), because Fe~{\sc ii} lines are so weakened that their 
measurements must suffer larger uncertainties. 

It may be worth comparing the spectroscopic parameters with those determined by other 
methods in recent representative studies. The comparisons with the results of Wang et al. 
(2009), Ram\'{\i}rez et al. (2013), and Luck (2017) are shown in  Figures~8, 9, and 10, 
respectively. In all three investigations, $T_{\rm eff}$ was determined 
photometrically from colors, $\log g$ by comparing the position on the $\log L$ vs. 
$\log T_{\rm eff}$ diagram ($L$: stellar luminosity) with stellar evolutionary tracks,
and $v_{\rm t}$ by requiring that the resulting abundances from Fe~{\sc i} lines do not 
show any systematic correlation with line strengths (though $v_{\rm t} = 1$~km~s$^{-1}$
was assumed by Wang et al. 2009). 
We can read the following characteristic trends from these figures.
\begin{itemize}
\item
Our spectroscopic $T_{\rm eff}$ is satisfactorily consistent with the photometrically
determined values of all three studies (Figures~8a, 9a, and 10a\footnote{
One exceptional disagreement is that our $T_{\rm eff}$ (4495~K) for $\xi$~Boo~B or HD~131156B
is considerably discrepant from Luck (2017)'s value (5240~K). We suspect that something 
was wrong in his derivation (e.g., adoption of the merged color of A+B?), because 
it is too high for a K5V star.}).
\item
Since the range of $\log g$ is rather narrow in G--K dwarfs (unlike the case of $T_{\rm eff}$),
our spectroscopic $\log g$ does not appear to be well correlated with the values based on the
theoretical HR diagram. However, the differences themselves are not so large, which are mostly within 
$\lesssim$~0.1--0.2~dex. We see on average that $\log g$(Wang) (Figure~8b) tends to be somewhat lower,  
while $\log g$(Ram\'{\i}rez) (Figure~9b) and $\log g$(Luck) (Figure~10b) somewhat higher, 
as compared with our $\log g$ derived from Fe~{\sc i} and Fe~{\sc ii} lines.
\item
Regarding $v_{\rm t}$, while Ram\'{\i}rez et al.'s (2013) results are almost consistent
with our determination (Figure~9c), those of Luck (2017) show some systematic trend
(Figure~10c) that they tend to be larger for higher $v_{\rm t}$ (while somewhat smaller 
for lower $v_{\rm t}$). We can see from Figure~8c that $v_{\rm t} =1$~km~s$^{-1}$ 
assumed by Wang et al. (2009) was not so a bad choice. 
\item
As to [Fe/H], a good agreement is confirmed with all these studies (cf. Figures~8d, 9d, 
and 10d).
\end{itemize}

As another check for the spectroscopic $T_{\rm eff}$ adopted in this study, 
we also computed the photometric $T_{\rm eff}$ from $B-V$ and [Fe/H] 
by using Casagrande et al.'s (2010) calibration based on the infrared 
flux method.\footnote{While their $T_{\rm eff}$--$(B-V)_{0}$--[Fe/H] 
relation was applied to our sample of 47 Hyades and 101 field stars, 
we also checked how the results are changed by using other color indices. 
Adopting Pinsonneault et al.'s (2004) $V-K_{s}$ and $J-K_{s}$ values for Hyades 
dwarfs (for which 37 stars at 4500~K~$\lesssim T_{\rm eff} \lesssim$~6300~K 
are in common with our sample), we determined $T_{\rm eff}^{V-K_{s}}$ and 
$T_{\rm eff}^{J-K_{s}}$ and compared them with $T_{\rm eff}^{B-V}$. 
The mean differences were found to be  
$\langle T_{\rm eff}^{V-K_{s}}-T_{\rm eff}^{B-V} \rangle = -6 (\pm 46)$~K and
$\langle T_{\rm eff}^{J-K_{s}}-T_{\rm eff}^{B-V} \rangle = +90 (\pm 94)$~K.
This suggests that, while Casagrande et al.'s (2010) calibration formula
results in quite consistent $T_{\rm eff}^{V-K_{s}}$ and $T_{\rm eff}^{B-V}$, 
it yields systematically higher $T_{\rm eff}^{J-K_{s}}$ than $T_{\rm eff}^{B-V}$ 
by $\sim 100$~K (with somewhat larger scatter).}
The comparisons between $T_{\rm eff}$(This study) 
and $T_{\rm eff}$(Casagrande) are shown in Figures 11a and 11b, 
where we can recongize that both are in satisfactory agreement.

It is also worthwhile to examine how our adopted spectroscopic 
$\log g_{\rm spec}$ is compared with the direct value ($\log g_{TLM}$) 
derived from $T_{\rm eff}$, $L$ (bolometric luminosity), and $M$ (mass).
The $L$ values were derived from $V$ (apparent magnitude; cf. Table~1), 
{\it Gaia} DR2 parallax (taken from the SIMBAD database; see also Section~3.2), 
and the bolometric correction evaluated by interpolating Alonso et al.'s (1995) Table~4.
Then, $M$ for each star was evaluated from its position on the 
$\log L$ vs. $\log T_{\rm eff}$ diagram (cf. Figure~11c) by comparing 
the theoretical PARSEC tracks (Bressan et al. 2012, 2013), where fine 
grids are available with a step of 0.005~$M_{\odot}$ over a wide metallicity 
range from $z = 0.0001$ to $z = 0.06$ (we regard 
$z = z_{\odot} \times 10^{{\rm [Fe/H]}}$ as the stellar metallicity
where $z_{\odot} = 0.014$). 
The difference between $\log g_{\rm spec}$ and the 
resulting $\log g_{TLM}$ is plotted against $T_{\rm eff}$ in Figure~11d,
which suggests that both are mostly consistent within $\sim \pm 0.1$~dex
(though several $\log g_{\rm spec}$ values are appreciably underestimated 
at $T_{\rm eff} \lesssim 5000$~K; see also Figure~2a).
These $\log L$, $M$, and $\log g_{TLM}$ values determined for each star 
are given in ``tableE1.dat'' of the online material.

\subsection{Oxygen Abundance from O~I 7771--5}

We go on to discussing the second subject of this study: whether or not credible 
oxygen abundances of K dwarfs can be derived from the high-excitation O~{\sc i} 
triplet lines at 7771--5~\AA, for which unreasonably high abundances were reported
by Schuler et al.'s group in their studies on open clusters (cf. Section~1).
As was the case for stellar parameters, Hyades G--K dwarfs can serve as an important 
touchstone in this respect, because they should retain almost the same (primordial) 
oxygen abundances in their photospheres.

Schuler et al. (2006a) derived  a markedly increasing [O/H](LTE) for Hyades dwarfs 
with a decrease in $T_{\rm eff}$; i.e., $\sim +0.2$ (at $T_{\rm eff} \sim$~6000--5500~K),
$\sim +0.5$ (at $T_{\rm eff} \sim 5000$~K), and $\sim +1.0$ (at $T_{\rm eff} \sim 4500$~K)
as shown in their Fig.~3. Their values are reproduced in Figure~12a (crosses) for 37 
stars in common with our sample. 
However, our results for Hyades stars turned out markedly different from theirs as 
manifestly seen from Figure~12a, where [O/H](NLTE) and [O/H](LTE) (represented by 
filled and open symbols, respectively) are plotted against $T_{\rm eff}$.\footnote{
The difference between [O/H](NLTE) and [O/H](LTE) (which is $\lesssim 0.1$~dex and 
quantitatively insignificant) changes sign around $T_{\rm eff} \sim 5800$~K, because 
non-LTE corrections both for the Sun and the star are involved in [O/H] 
($\equiv A_{\rm star}({\rm O}) - A_{\odot}(\rm O)$).}
That is, our [O/H] values do not show any such progressive increase towards 
lower $T_{\rm eff}$ as reported by Schuler et al. (2006a) but distribute around $\sim +0.2$,
being consistent with the expectation that these stars should show similar oxygen   
abundances.

We investigated the cause of this discrepancy by comparing the adopted stellar parameters
in both studies. Comparisons of $T_{\rm eff}$, $\log g$, and $v_{\rm t}$ are  
illustrated in Figures~12b, 12c, and 12d, respectively.
It is apparent from Figure~12b that a considerable disagreement exists between 
Schuler et al.'s $T_{\rm eff}$ (photometric determination using colors) and our 
$T_{\rm eff}$ (spectroscopic determination from Fe lines) in the sense that the former is
systematically lower by several hundred K and the difference progressively increasing
towards lower $T_{\rm eff}$. Meanwhile, a more or less reasonable consistency (excepting 
an outlier) is observed for $\log g$ (Figure~12c), which they derived from theoretical
evolutionary tracks. As to $v_{\rm t}$,
Schuler et al.'s values tend to be somewhat higher than ours especially in the regime of 
larger $v_{\rm t}$ or higher $T_{\rm eff}$ (Figure~12d). This disagreement may be explained 
by the fact that they used Allende Prieto et al.'s (2004) empirical formula derived for 
field stars and that our $v_{\rm t}$ values derived for Hyades dwarfs tends to be lower 
than those of field dwarfs at $T_{\rm eff} \gtrsim 5500$~K as remarked in Section~3.2 
(cf. Figure~2b). 

In view of these results along with the parameter-dependence of the abundances discussed in Section~4.2, 
it must be the difference in $T_{\rm eff}$ that is mainly responsible for the 
considerable discrepancy in [O/H] between Schuler et al. (2006a) and this study, 
because the oxygen abundance from high-excitation O~{\sc i} 7771--5 triplet is highly 
sensitive to a change in $T_{\rm eff}$ (Figure~6e) while the roles played by $\log g$ and 
$v_{\rm t}$ are insignificant (Figures~6f and 6g). 
This can be confirmed from Figure~12e, where 
$\chi_{\rm low}\theta_{\rm eff}$(Schuler)$-$$\chi_{\rm low}\theta_{\rm eff}$(This study)
($\theta_{\rm eff} \equiv 5040/T_{\rm eff}$; this is the expected abundance variation
for neutral oxygen of dominant population due to the difference in $T_{\rm eff}$) is 
plotted against $T_{\rm eff}$ for each star. We can see from this figure that
the abundance change systematically grows with a decrease in $T_{\rm eff}$ 
(from $\sim$~0.1--0.2~dex at $T_{\rm eff} \sim 6000$~K up to $\sim 0.6$~dex at 
$T_{\rm eff} \sim 4500$~K), which reasonably explains why Schuler et al.'s
[O/H] values tend to be progressively larger than ours towards lower $T_{\rm eff}$.
Besides, we found that the equivalent widths of the O~{\sc i} triplet lines measured 
by them and used for their analysis tend to be somewhat overestimated (by several tens \%) 
for weak lines ($W_{\lambda} \lesssim 20~$m\AA) in comparison with our values (Figure~12f), 
though both are consistent for lines of medium strength. 
This would have further enhanced the overestimation of their [O/H] in case of such small 
$W_{\lambda}$ (i.e., $T_{\rm eff} \lesssim 5000$~K). 
As such, we consider that Schuler et al.'s (2006a) anomalous [O/H] results derived for Hyades
dwarfs (conspicuously increasing towards lower $T_{\rm eff}$) are mainly due to their 
inadequate $T_{\rm eff}$ scale (i.e., too low by several hundred K) and thus should not 
be seriously taken.

The results of this study suggest that consistent oxygen abundances for Hyades G--K dwarfs 
(i.e., without showing any systematic trend in terms of $T_{\rm eff}$) can be derived 
even based on the high-excitation O~{\sc i} 7771--5 triplet lines.
The mean non-LTE $\langle$[O/H]$\rangle$ for 37 stars (common to Schuler et al.) 
depicted in Figure~12a is $+0.12$ ($\sigma = 0.09$), and that for all our 47 Hyades 
stars (cf. Figure~6c) is identically $+0.12$ ($\sigma = 0.09$), which are favorably compared
(i.e., within error bars) with $\langle$[O/H]$\rangle = +0.22 (\pm 0.14)$ obtained by 
Takeda et al. (2013) for Hyades F--G stars based on O~{\sc i} 6156--8 lines.   

Even so, it should be kept in mind that precision of abundance determination would 
naturally deteriorate for K dwarfs ($T_{\rm eff} \lesssim 5000$~K) because the strengths of 
these high-excitation O~{\sc i} triplet lines are considerably weakened, which makes 
measurement more difficult (e.g. due to increased importance of blending by other lines). 
This can be manifestly seen from the appreciable scatter of [O/H] at 
$T_{\rm eff} \lesssim 5000$~K in Figure~12a.
Yet, we would like to stress that such significant ``$T_{\rm eff}$-dependent 
systematic trend'' as reported by Schuler et al. (2006a) is unlikely. 

Admittedly, what has been argued above is specific to Hyades dwarfs and we can not say 
much about the $T_{\rm eff}$-dependent anomaly in [O/H] derived from O~{\sc i} 7771--5 
(i.e., progressively increasing towards lower $T_{\rm eff}$) also reported for other 
cluster stars: e.g., UMa moving group (King \& Schuler 2005); Pleiades and M~34 
(Schuler et al. 2004); NGC~752 (Maderak et al. 2013). We consider, however, 
that almost the same argument may also apply to the consequences of these studies, 
because we confirmed that the $T_{\rm eff}$ scale they adopted tends to be 
systematically lower as compared with that of Casagrande et al. (2010)
(which is consistent with our spectrooscopic $T_{\rm eff}$; cf. Figures~11a and 11b).
The details of this examination are separately described in Appendix~A (see also 
Table~2 and Figure~14).
 
Turning our attention to field stars, we compare our oxygen abundances with those 
derived by three previous studies (as done in Section~5.1 for stellar parameters).
The panels (e) and (f) of Figures 8, 9, and 10 show comparison of our (non-LTE) 
[O/H] and [O/Fe] values with those of Wang et al. (2009: from O~{\sc i} 7771--5 with non-LTE), 
Ram\'{\i}rez et al. (2013; from O~{\sc i} 7771-5 with non-LTE), and Luck (2017; 
from [O~{\sc i}] 6300 with LTE), respectively. These figures suggest that rough 
correlation is observed (though not necessarily good) between our and their results.
In addition, in Figure~13 is compared the non-LTE [O/Fe] vs. [Fe/H] relation derived 
in this study for 148 G--K dwarfs (47 Hyades stars at 6300~K~$\gtrsim T_{\rm eff} \gtrsim$~4500~K 
and 101 field stars at 5500~K~$\gtrsim T_{\rm eff} \gtrsim$~4500~K) with
the similar relation obtained by Takeda \& Honda (2005) based on the same O~{\sc i} 
7771--5 triplet (with non-LTE) for early F--early K dwarfs/subgiants 
(at 7000~K~$\gtrsim T_{\rm eff} \gtrsim$~5000~K).
It can be confirmed by comparing panels (a) and (b) of Figure~13 that quite a similar 
trend of [O/Fe] (i.e., increasing with a decrease in [Fe/H] with almost the same gradient) 
is observed for both cases. This is a reasonable consequence, because most of the sample 
stars belong to the thin-disk population in this study (cf. Section~3.2)
as well as in Takeda \& Honda (2005) (cf. Sect.~2.2 in Takeda 2007).
For comparison, the similar relations between [O/Fe] and [Fe/H] derived by Hawkins et al. (2016)
for a large number of disk stars (APOGEE+Kepler sample) are overplotted in these figures.
Although the global tendency of decreasing [O/Fe] with an increase in [Fe/H] is similar,
their [O/Fe] tends to be stagnant and supersolar (i.e., $\gtrsim 0$) at [Fe/H]~$\gtrsim 0$
unlike our results ([O/Fe]~$\lesssim 0$ at at [Fe/H]~$\gtrsim 0$). See also Sect.~4.1
in Hawkins et al. (2016).

Combining all the results mentioned above, we may conclude
that oxygen abundances can be reliably determined based on the O~{\sc i} triplet lines 
at 7771--5~\AA\ for K dwarfs (just like F and G stars), as long as stars
of $T_{\rm eff} \gtrsim 4500$~K are concerned (actually, $\sim 4500$~K corresponding to
spectral type of $\sim$~K~5 is the practical lower limit of $T_{\rm eff}$, below which 
these high-excitation O~{\sc i} lines become too weak to be usable).  

\section{SUMMARY AND CONCLUSION}

It has been reported that Fe abundances of K dwarfs derived from 
Fe~{\sc i} and Fe~{\sc ii} lines tend to show considerable discrepancy 
(i.e., the latter is larger than the former), becoming progressively 
more serious with a decrease in $T_{\rm eff}$.
If it is real, the widely used spectroscopic method for determining 
the parameters of solar-type stars based on Fe lines (which makes use 
of ionization equilibrium of Fe~{\sc i}/Fe~{\sc ii}) would hardly 
be applicable to K dwarfs, since classical model atmospheres would 
be no more valid for them.

According to the recent investigations of Aleo et al. (2017) and
Tsantaki ey al. (2019), however, the alleged large Fe~{\sc ii}--Fe~{\sc i} 
disagreement in K dwarfs is likely to be due to the use of blending-affected 
Fe~{\sc ii} lines, and can be appreciably mitigated down to a practically 
insignificant level when they are removed.
This may suggest the necessity of reexamining another similar problem
related to K dwarfs (argued by Schuler et al. from their studies on 
open cluster stars) that oxygen abundances derived from the high-excitation 
O~{\sc i} 7771--5 triplet lines are strikingly overestimated (even by up 
to $\sim 1$~dex), its extent becoming more prominent towards lower $T_{\rm eff}$.

Motivated by this situation, we decided to reexamine whether these 
``spectroscopic K dwarf problems'' really exist, based on the spectral 
data of 148 G--K dwarfs (47 Hyades stars and 101 field stars). 
This may be checked by applying the conventional method of analysis 
(for determining stellar parameters and oxygen abundances) to these 
program stars. That is, some kind of unreasonable or inconsistent result 
must be observed if the classical modeling really breaks down for K dwarfs. 

We determined $T_{\rm eff}$, $\log g$, $v_{\rm t}$, and [Fe/H] for all the 
program stars based on the equivalent widths of Fe~{\sc i} and Fe~{\sc ii} 
lines as done by Takeda et al. (2005).
Comparing our spectroscopic $T_{\rm eff}$ and $\log g$ of Hyades stars with 
those of de Bruijne et al. (2001) (which are considered to be well established),
we found that the differences are practically not so significant (especially, 
no evidence was found that K dwarfs suffer larger errors than G dwarfs).
This result may support Aleo et al.'s (2017) conclusion that the differences 
between Fe~{\sc i} and Fe~{\sc ii} abundances in K dwarfs are actually not 
so important ($\lesssim 0.1$~dex) at least for stars of $T_{\rm eff} \gtrsim 4500$~K. 

The oxygen abundances of these G--K dwarfs were derived by applying the
spectrum-fitting technique to the 7770--7782~\AA\ region comprising
O~{\sc i} 7771--5 and Fe~{\sc i} 7780 lines, where the non-LTE effect
was taking into account for the O~{\sc i} lines.
Regarding the [O/H] values of Hyades stars, our results turned out to
distribute around $\sim +0.2$ (being consistent with the expectation that 
cluster stars should have similar abundances), in marked contrast with the
progressively increasing tendency (even up to $\sim +1$) towards lower $T_{\rm eff}$ 
reported by Schuler et al. (2006a).

We investigated the reason for this distinct discrepancy, and found that
the higher $T_{\rm eff}$ scale adopted by them is the main cause,
which has a large impact on the abundances derived from O~{\sc i} lines 
of high-excitation. It was also confirmed that the [O/Fe] vs. [Fe/H] relation 
we obtained for 101 field mid G--mid K stars is quite similar to that derived 
by Takeda \& Honda (2005) for 160 stars (mainly F--G type), which means 
that K dwarfs can not be exceptionally anomalous in terms of oxygen abundance
determination based on the O~{\sc i} 7771--5 triplet.

In summary, we conclude for K dwarfs that their atmospheric parameters can be 
spectroscopically evaluated to a sufficient precision in the conventional manner 
using Fe lines (because the classical Saha--Boltzmann equation for Fe~{\sc i}/Fe~{\sc ii} 
is still not a bad assumption) and oxygen abundances can be reliably established from the 
high-excitation O~{\sc i} 7771--5 triplet (just like F--G dwarfs), so far as stars 
of $T_{\rm eff} \gtrsim 4500$~K are concerned. 

\bigskip

This investigation is based in part on the data collected at Subaru Telescope, 
which is operated by the National Astronomical Observatory of Japan.
Data reduction was in part carried out by using the common-use data analysis 
computer system at the Astronomy Data Center (ADC) of the National Astronomical 
Observatory of Japan.
This research has made use of the SIMBAD database, operated by CDS, 
Strasbourg, France. 
This work also used the data from the European Space Agency (ESA) mission {\it Gaia}, 
processed by the {\it Gaia} Data Processing and Analysis Consortium (DPAC). 
Funding for the DPAC has been provided by national institutions, in particular 
the institutions participating in the {\it Gaia} Multilateral Agreement.

\newpage

\appendix

\section{IMPACT OF EFFECTIVE TEMPERATURE SCALE ON [O/H] IN PREVIOUS STUDIES OF OPEN CLUSTERS}

We showed in Section~5.2 that the conspicuous excess of [O/H] increasing toward 
a lower $T_{\rm eff}$ concluded by Schuler et al. (2006a) for Hyades stars
could be interpreted as mainly due to the systematically lower $T_{\rm eff}$ scale
they adopted (cf. Figure~12). 
Regarding the similar tendencies in [O/H] (based on the high-excitation 
O~{\sc i} 7771--5 triplet lines) also reported by several authors for open clusters 
other than Hyades (i.e., UMa moving group, M~34, Pleiades, NGC~752; cf. Section~1),
we are unable to check them directly as done in Figure~12.
Still, however, we can examine whether the $T_{\rm eff}$ scales adopted by those 
previous studies are reasonable and how they affect the [O/H] trends.

We first postulate that Casagrande et al.'s (2010) calibration yields reasonably
correct $T_{\rm eff}$, which we confirmed to be consistent with our spectroscopic 
$T_{\rm eff}$ (cf. Figures~11a and 11b). Since $T_{\rm eff}$ values were derived
photometrically from $B-V$ colors by using any of the following three formulas 
in most of these relevant studies (cf. Table~2 for a brief summary), the effect 
we want to examine is essentially attributed to the difference of these equations 
from that of Casagrande et al. (2010). 
\begin{equation}
\begin{split}
T_{\rm eff} & = 5040 / \bigl(0.5247+0.5396 (B-V)_{0}\bigr) \\
            & + 701.7(B-V)_{0} \bigl([{\rm Fe/H}] - [{\rm Fe/H}]_{\rm Hyades}\bigr),
\end{split}
\end{equation} 
\begin{equation}
T_{\rm eff} = 1808 (B-V)_{0}^{2} - 6103 (B-V)_{0} + 8899,
\end{equation}
and
\begin{equation}
\begin{split}
T_{\rm eff} & = 8344 - 3631.32 (B-V)_{0} - 2211.81 (B-V)_{0}^{2} \\ 
            & + 3365.44 (B-V)_{0}^{3} - 1033.62 (B-V)_{0}^{4} \\
            & +701.7 \bigl([{\rm Fe/H}] - [{\rm Fe/H}]_{\rm Hyades}\bigr),
\end{split}
\end{equation} 
where $T_{\rm eff}$ is in K, $(B-V)_{0}$ is the reddening-corrected $B-V$ color, [Fe/H] is 
the metallicity of a star, and [Fe/H]$_{\rm Hyades}$ is the Hyades metallicity (assumed to be
0.15 in this study).
These $T_{\rm eff}$ vs. $(B-V)_{0}$ relations of Equations (A1), (A2), and (A3) are 
compared with that of Casagrande et al. (2010) in Figure~14a, where we can see that 
all the former three tend to yield systematically lower $T_{\rm eff}$ than the latter 
at $(B-V)_{0} \gtrsim 0.6$ with discrepancies increasing towards lower $T_{\rm eff}$.

For each star, $T_{\rm eff}^{\rm Casagrande}$ was computed from $(B-V)_{0}$ (taken from 
the relevant paper) and assumed cluster [Fe/H] (cf. Table~2) according to Casagrande 
et al.'s (2010) recipe and compared with literature value ($T_{\rm eff}^{\rm literature}$) 
actually adopted therein. The differences  
$\chi_{\rm low}\theta_{\rm eff}^{\rm literature} - \chi_{\rm low}\theta_{\rm eff}^{\rm Casagrande}$
(see the caption of Figure~12e for the meanings of $\theta_{\rm eff}$ and $\chi_{\rm low}$)
are plotted against $T_{\rm eff}^{\rm Casagrande}$ in Figure~14b, 
from which we can read the following characteristics.
\begin{itemize}
\item
In all cases, the differences between $\chi_{\rm low}\theta_{\rm eff}^{\rm literature}$ and 
$\chi_{\rm low}\theta_{\rm eff}^{\rm Casagrande}$, which correspond to the expected overestimation
of [O/H] due to an underestimated $T_{\rm eff}$ (cf. Section 5.2), tend to progressively increase 
with a lowering of $T_{\rm eff}$; i.e., from $\sim 0.0$~dex (at $\sim 6000$~K)
to $\sim$~0.3--0.5~dex (at $\sim 5000$~K). This reasonably explains the $T_{\rm eff}$-dependent 
tendency of [O/H] concluded in their papers (cf. Table~2) at least qualitatively,
which indicates that the inappropriate $T_{\rm eff}$ scale is the main cause for the trend.  
\item
Quantitatively, however, only this $T_{\rm eff}$-related correction seems to be
rather insufficient to account for the differences ([O/H]$_{5000}-$[O/H]$_{6000}$) ranging
from $\sim$~0.2~dex to $\sim$~0.7~dex (Table~2), which means that some other factors 
(such as an underestimation of $W_{\lambda}$ for the very weak line case at the low-$T_{\rm eff}$ 
regime; cf. Figure~12f) may also be involved.    
\item
Especially, as seen from Fig.~2 of Schuler et al. (2004), the d[O/H]/d$T_{\rm eff}$ 
gradient of Pleiades and M~34 cluster stars at $T_{\rm eff} \lesssim$~5200~K appears to
become abruptly steeper. Since these two open clusters are younger than the Hyades and thus 
stellar activity should be higher, a possibility may not be excluded that some activity-related
effect might influence the strength of high-excitation O~{\sc i} triplet for these cases. 
Accordingly, oxygen abundances from O~{\sc i} 7771--5 lines for dwarfs of these younger clusters
at the $T_{\rm eff}$ regime of 4500~K~$\lesssim T_{\rm eff} \lesssim 5500$~K may be worth 
careful reinvestigation based on reliable $T_{\rm eff}$ scales. 
\end{itemize}

\clearpage

\onecolumn

\setcounter{table}{0}
\begin{table}[h]
\scriptsize
\caption{Basic data, parameters, and abundance results of 148 program stars.}
\begin{center}
\begin{tabular}{rrrrrrrrrrrrr
}\hline\hline                 
HIP & HD & $V$ & $M_{V}$ & $B-V$ & $T_{\rm eff}$ & $\log g$ & $v_{\rm t}$ & [Fe/H] &  
 $v_{\rm M}$ & [O/H] & $\Delta_{7774}$ & $W_{7774}$ \\
(1) & (2) & (3) & (4) & (5) & (6) & (7) & (8) & (9) & (10) &(11) &(12) &(13) \\
\hline
\multicolumn{13}{c}{(Hyades stars)}\\
 19796 &  26784 & 7.11 & 3.73 & 0.51 & 6307 & 4.29 & 1.37 & 0.26& 13.74 &  0.19 & $-$0.23& 125.4\\
 22566 &  30809 & 7.90 & 4.07 & 0.53 & 6250 & 4.29 & 1.10 & 0.24&  9.06 &  0.14 & $-$0.21& 113.1\\
 19386 &  26257 & 7.64 & 3.57 & 0.55 & 6201 & 4.26 & 1.03 & 0.23&  6.31 &  0.12 & $-$0.21& 108.2\\
 20557 &  27808 & 7.13 & 4.07 & 0.52 & 6199 & 4.27 & 1.21 & 0.19& 10.45 &  0.17 & $-$0.21& 114.9\\
 20815 &  28205 & 7.41 & 4.11 & 0.54 & 6180 & 4.35 & 1.07 & 0.19&  8.67 &  0.22 & $-$0.20& 113.2\\
 25639 &  35768 & 8.50 & 3.82 & 0.56 & 6138 & 4.08 & 1.13 & 0.15&  5.83 &  0.06 & $-$0.23& 105.8\\
 20237 &  27406 & 7.46 & 4.20 & 0.56 & 6134 & 4.37 & 1.07 & 0.27&  9.59 &  0.19 & $-$0.18& 106.8\\
 15304 &  20430 & 7.38 & 3.91 & 0.57 & 6125 & 4.29 & 0.99 & 0.33&  5.97 &  0.17 & $-$0.18& 104.5\\
 10672 &  14127 & 8.55 & 4.48 & 0.57 & 6125 & 4.41 & 0.88 & 0.01&  6.58 & $-$0.05 & $-$0.16&  84.2\\
 22422 &  30589 & 7.72 & 4.19 & 0.58 & 6081 & 4.39 & 0.97 & 0.24&  5.54 &  0.07 & $-$0.15&  90.9\\
 19148 &  25825 & 7.85 & 4.50 & 0.59 & 6078 & 4.47 & 1.05 & 0.21&  6.14 &  0.06 & $-$0.14&  87.8\\
 15310 &  20439 & 7.78 & 4.46 & 0.62 & 6003 & 4.48 & 1.01 & 0.31&  5.99 &  0.16 & $-$0.14&  89.3\\
 20577 &  27859 & 7.79 & 4.37 & 0.60 & 5955 & 4.30 & 0.96 & 0.14&  6.44 &  0.11 & $-$0.16&  87.5\\
 21317 &  28992 & 7.90 & 4.73 & 0.63 & 5928 & 4.48 & 0.98 & 0.20&  5.55 &  0.13 & $-$0.12&  81.5\\
 19786 &  26767 & 8.05 & 4.78 & 0.64 & 5922 & 4.42 & 1.07 & 0.23&  5.60 &  0.08 & $-$0.12&  79.4\\
 20899 &  28344 & 7.83 & 4.45 & 0.61 & 5902 & 4.30 & 1.11 & 0.15&  6.68 &  0.15 & $-$0.15&  86.9\\
 20741 &  28099 & 8.10 & 4.75 & 0.66 & 5852 & 4.55 & 0.92 & 0.24&  4.42 &  0.12 & $-$0.10&  72.6\\
 19793 &  26736 & 8.05 & 4.73 & 0.66 & 5815 & 4.38 & 0.99 & 0.22&  5.78 &  0.12 & $-$0.12&  74.6\\
 19781 &  26756 & 8.45 & 5.15 & 0.69 & 5745 & 4.54 & 0.91 & 0.22&  5.13 &  0.08 & $-$0.09&  62.7\\
 20146 &  27282 & 8.47 & 5.11 & 0.72 & 5677 & 4.48 & 0.93 & 0.24&  5.32 &  0.09 & $-$0.09&  60.0\\
 23750 & 240648 & 8.82 & 5.19 & 0.73 & 5630 & 4.57 & 0.89 & 0.25&  5.38 &  0.12 & $-$0.07&  56.5\\
 14976 &  19902 & 8.15 & 5.03 & 0.73 & 5614 & 4.57 & 0.91 & 0.18&  3.72 &  0.08 & $-$0.07&  53.7\\
 20130 &  27250 & 8.62 & 5.48 & 0.74 & 5591 & 4.55 & 1.00 & 0.18&  4.59 &  0.08 & $-$0.07&  52.6\\
 23069 &  31609 & 8.89 & 5.36 & 0.74 & 5583 & 4.53 & 0.84 & 0.21&  3.87 &  0.10 & $-$0.07&  53.1\\
 24923 & 242780 & 9.03 & 5.34 & 0.77 & 5560 & 4.57 & 0.85 & 0.27&  5.04 &  0.12 & $-$0.07&  52.0\\
 21099 &  28593 & 8.59 & 5.28 & 0.73 & 5557 & 4.44 & 0.99 & 0.17&  4.68 &  0.07 & $-$0.08&  52.2\\
 23498 &  32347 & 9.00 & 5.33 & 0.77 & 5549 & 4.52 & 1.03 & 0.21&  4.96 &  0.05 & $-$0.07&  49.2\\
 20949 & 283704 & 9.19 & 5.35 & 0.77 & 5544 & 4.57 & 0.88 & 0.23&  3.75 &  0.09 & $-$0.07&  49.3\\
 20480 &  27732 & 8.84 & 5.41 & 0.76 & 5539 & 4.49 & 1.00 & 0.15&  4.55 &  0.11 & $-$0.07&  52.2\\
 21741 & 284574 & 9.40 & 5.42 & 0.81 & 5425 & 4.55 & 0.95 & 0.23&  5.00 &  0.17 & $-$0.06&  46.2\\
 19934 & 284253 & 9.14 & 5.59 & 0.81 & 5376 & 4.57 & 0.87 & 0.17&  3.71 &  0.15 & $-$0.05&  41.9\\
 20951 & 285773 & 8.95 & 5.87 & 0.83 & 5350 & 4.57 & 1.02 & 0.14&  3.90 &  0.12 & $-$0.05&  39.6\\
 22380 &  30505 & 8.98 & 5.63 & 0.83 & 5336 & 4.57 & 0.94 & 0.21&  4.49 &  0.19 & $-$0.05&  41.1\\
 20850 &  28258 & 9.02 & 5.66 & 0.84 & 5321 & 4.50 & 1.04 & 0.18&  4.19 &  0.15 & $-$0.05&  39.9\\
 20492 &  27771 & 9.11 & 5.74 & 0.85 & 5292 & 4.60 & 0.89 & 0.22&  4.38 &  0.01 & $-$0.04&  31.2\\
 20978 &  28462 & 9.08 & 6.04 & 0.86 & 5242 & 4.50 & 1.04 & 0.16&  4.74 &  0.09 & $-$0.04&  33.3\\
 18327 & 285252 & 8.99 & 5.91 & 0.90 & 5183 & 4.61 & 1.02 & 0.21&  4.27 &  0.18 & $-$0.04&  31.9\\
 19098 & 285367 & 9.31 & 5.79 & 0.89 & 5123 & 4.56 & 1.03 & 0.11&  4.41 &  0.24 & $-$0.04&  31.6\\
 23312 &  $\cdots$   & 9.71 & 5.83 & 0.96 & 5104 & 4.53 & 0.90 & 0.14&  4.00 &  0.22 & $-$0.04&  30.1\\
 20827 & 285830 & 9.48 & 5.67 & 0.93 & 5089 & 4.61 & 0.89 & 0.23&  3.88 &  0.22 & $-$0.03&  28.2\\
 20082 & 285690 & 9.57 & 6.08 & 0.98 & 5030 & 4.53 & 0.79 & 0.15&  3.60 &  0.01 & $-$0.03&  21.7\\
 19263 & 285482 & 9.94 & 6.41 & 1.00 & 4898 & 4.50 & 0.85 & 0.07&  4.24 &  0.03 & $-$0.02&  17.0\\
 22654 & 284930 &10.29 & 6.68 & 1.07 & 4831 & 4.47 & 0.99 & 0.11&  4.22 & $-$0.07 & $-$0.02&  12.5\\
 18946 &  $\cdots$   &10.12 & 6.94 & 1.09 & 4823 & 4.64 & 0.92 & 0.16&  3.46 & $-$0.05 & $-$0.01&  11.5\\
 20762 & 286789 &10.48 & 7.18 & 1.15 & 4729 & 4.25 & 1.15 &$-$0.02&  4.51 & $-$0.20 & $-$0.02&   9.0\\
 18322 & 286363 &10.12 & 7.24 & 1.07 & 4725 & 4.71 & 0.91 & 0.17&  3.97 &  0.31 & $-$0.01&  14.2\\
 19441 &  $\cdots$   &10.10 & 7.47 & 1.19 & 4525 & 4.61 & 0.70 & 0.14&  3.56 &  0.11 & $-$0.01&   6.3\\
\hline
\multicolumn{13}{c}{(Field stars)}\\
053471 &  94718 & 8.45 & 5.58 & 0.73 & 5482 & 4.45 & 0.73 &$-$0.03&  3.47 & $-$0.13 & $-$0.06&  37.2\\
082588 & 152391 & 6.65 & 5.51 & 0.75 & 5475 & 4.50 & 0.94 & 0.02&  4.15 & $-$0.01 & $-$0.06&  42.0\\
004907 &   5996 & 7.67 & 5.61 & 0.76 & 5445 & 4.53 & 0.96 &$-$0.09&  3.68 & $-$0.05 & $-$0.06&  38.6\\
010818 &  14374 & 8.48 & 5.51 & 0.74 & 5444 & 4.58 & 0.80 &$-$0.01&  3.42 & $-$0.02 & $-$0.05&  38.4\\
026653 &  37216 & 7.85 & 5.63 & 0.76 & 5441 & 4.58 & 0.94 &$-$0.03&  3.48 & $-$0.02 & $-$0.05&  38.5\\
059280 & 105631 & 7.46 & 5.53 & 0.79 & 5439 & 4.50 & 0.88 & 0.19&  3.57 &  0.01 & $-$0.06&  40.5\\
040419 &  69076 & 8.27 & 5.61 & 0.71 & 5434 & 4.47 & 0.66 &$-$0.24&  3.47 & $-$0.23 & $-$0.05&  31.3\\
093926 & 178450 & 7.78 & 5.54 & 0.76 & 5423 & 4.45 & 2.00 &$-$0.04& 18.51 &  0.24 & $-$0.07&  57.4\\
\hline
\end{tabular}
\end{center}
\end{table}

\setcounter{table}{0}
\begin{table}[h]
\scriptsize
\caption{(Continued.)}
\begin{center}
\begin{tabular}{rrrrrrrrrrrrr
}\hline\hline                 
HIP & HD & $V$ & $M_{V}$ & $B-V$ & $T_{\rm eff}$ & $\log g$ & $v_{\rm t}$ & [Fe/H] &  
 $v_{\rm M}$ & [O/H] & $\Delta_{7774}$ & $W_{7774}$ \\
(1) & (2) & (3) & (4) & (5) & (6) & (7) & (8) & (9) & (10) &(11) &(12) &(13) \\
\hline
043852 &  76218 & 7.69 & 5.60 & 0.77 & 5422 & 4.59 & 0.83 & 0.03&  4.34 & $-$0.04 & $-$0.05&  35.9\\
094346 & 180161 & 7.04 & 5.53 & 0.80 & 5418 & 4.49 & 0.95 & 0.17&  3.67 &  0.11 & $-$0.06&  44.2\\
055210 &  98281 & 7.29 & 5.58 & 0.73 & 5401 & 4.47 & 0.75 &$-$0.21&  3.24 & $-$0.15 & $-$0.05&  33.1\\
112245 & 215500 & 7.50 & 5.50 & 0.72 & 5399 & 4.39 & 0.62 &$-$0.20&  3.15 & $-$0.15 & $-$0.06&  33.8\\
046843 &  82443 & 7.05 & 5.80 & 0.78 & 5393 & 4.65 & 1.27 &$-$0.03&  5.38 &  0.07 & $-$0.05&  39.5\\
098677 & 190067 & 7.15 & 5.72 & 0.71 & 5376 & 4.48 & 0.64 &$-$0.32&  3.23 & $-$0.37 & $-$0.04&  24.4\\
065515 & 116956 & 7.29 & 5.59 & 0.80 & 5372 & 4.51 & 1.07 & 0.14&  5.30 &  0.11 & $-$0.05&  41.5\\
115331 & 220182 & 7.36 & 5.66 & 0.80 & 5368 & 4.56 & 1.08 & 0.06&  4.98 &  0.01 & $-$0.05&  36.3\\
007576 &  10008 & 7.66 & 5.79 & 0.80 & 5358 & 4.52 & 0.97 &$-$0.03&  3.48 & $-$0.11 & $-$0.04&  31.6\\
062016 & 110514 & 8.04 & 5.61 & 0.80 & 5358 & 4.49 & 0.73 &$-$0.01&  3.40 & $-$0.01 & $-$0.05&  35.4\\
075277 & 136923 & 7.16 & 5.64 & 0.80 & 5357 & 4.56 & 0.67 &$-$0.05&  3.35 & $-$0.06 & $-$0.04&  32.4\\
010798 &  14412 & 6.33 & 5.81 & 0.72 & 5357 & 4.51 & 0.73 &$-$0.49&  3.30 & $-$0.35 & $-$0.04&  24.9\\
010276 &  13483 & 8.46 & 5.81 & 0.78 & 5347 & 4.54 & 0.85 &$-$0.17&  3.41 & $-$0.08 & $-$0.05&  31.9\\
042074 &  72760 & 7.32 & 5.63 & 0.79 & 5344 & 4.59 & 0.92 & 0.09&  3.64 &  0.07 & $-$0.04&  36.4\\
050782 &  89813 & 7.78 & 5.64 & 0.75 & 5336 & 4.51 & 0.67 &$-$0.07&  3.28 & $-$0.11 & $-$0.04&  30.3\\
081813 & 151541 & 7.56 & 5.63 & 0.77 & 5334 & 4.42 & 0.66 &$-$0.14&  3.40 & $-$0.17 & $-$0.05&  29.6\\
028954 &  41593 & 6.76 & 5.81 & 0.81 & 5332 & 4.50 & 1.04 & 0.05&  4.54 &  0.00 & $-$0.05&  34.7\\
106122 & 204814 & 7.93 & 5.56 & 0.76 & 5327 & 4.44 & 0.66 &$-$0.20&  3.18 &  0.09 & $-$0.06&  39.4\\
000400 & 225261 & 7.82 & 5.78 & 0.76 & 5323 & 4.49 & 0.62 &$-$0.38&  3.25 & $-$0.12 & $-$0.05&  30.3\\
077408 & 141272 & 7.44 & 5.79 & 0.80 & 5304 & 4.45 & 1.01 & 0.02&  4.20 & $-$0.07 & $-$0.05&  31.3\\
051819 &  90343 & 7.29 & 5.68 & 0.82 & 5303 & 4.50 & 0.78 & 0.12&  3.50 &  0.04 & $-$0.05&  34.2\\
014023 &  18702 & 8.11 & 5.56 & 0.84 & 5280 & 4.47 & 0.74 & 0.18&  3.09 &  0.21 & $-$0.05&  40.0\\
085235 & 158633 & 6.44 & 5.90 & 0.76 & 5270 & 4.54 & 0.60 &$-$0.41&  3.05 & $-$0.26 & $-$0.04&  24.5\\
074702 & 135599 & 6.92 & 5.96 & 0.83 & 5250 & 4.63 & 0.90 &$-$0.04&  4.05 & $-$0.01 & $-$0.04&  28.6\\
116085 & 221354 & 6.76 & 5.63 & 0.84 & 5246 & 4.53 & 0.64 & 0.10&  3.01 &  0.17 & $-$0.04&  35.6\\
072200 & 130215 & 7.98 & 5.87 & 0.87 & 5244 & 4.56 & 0.88 & 0.13&  3.42 &  0.05 & $-$0.04&  31.0\\
039157 &  65583 & 6.97 & 5.84 & 0.72 & 5243 & 4.54 & 0.53 &$-$0.71&  3.00 & $-$0.11 & $-$0.05&  27.3\\
082267 & 151877 & 8.40 & 5.85 & 0.82 & 5237 & 4.59 & 0.69 &$-$0.10&  3.04 & $-$0.09 & $-$0.04&  25.9\\
002742 &   3141 & 8.02 & 5.71 & 0.87 & 5225 & 4.51 & 0.69 & 0.18&  3.04 &  0.13 & $-$0.04&  32.9\\
012926 &  17190 & 7.89 & 5.84 & 0.84 & 5224 & 4.61 & 0.61 &$-$0.05&  3.10 & $-$0.03 & $-$0.04&  26.8\\
033848 &  52456 & 8.16 & 5.90 & 0.86 & 5212 & 4.51 & 0.69 & 0.06&  3.25 & $-$0.02 & $-$0.04&  27.6\\
078241 & 143291 & 8.02 & 5.94 & 0.76 & 5208 & 4.40 & 0.50 &$-$0.40&  3.22 & $-$0.20 & $-$0.04&  24.0\\
039064 &  65430 & 7.68 & 5.86 & 0.83 & 5202 & 4.55 & 0.57 &$-$0.09&  2.99 &  0.09 & $-$0.04&  30.5\\
008543 &  11130 & 8.06 & 5.92 & 0.76 & 5197 & 4.52 & 0.52 &$-$0.57&  2.92 & $-$0.13 & $-$0.04&  24.7\\
000184 & 224983 & 8.48 & 5.85 & 0.89 & 5195 & 4.55 & 0.73 & 0.13&  2.94 &  0.02 & $-$0.04&  27.5\\
061099 & 108984 & 7.91 & 5.90 & 0.86 & 5194 & 4.54 & 0.60 & 0.11&  3.14 &  0.09 & $-$0.04&  29.6\\
010532 &  13977 & 9.11 & 5.79 & 0.88 & 5188 & 4.58 & 0.72 & 0.11&  3.20 &  0.06 & $-$0.04&  28.2\\
066781 & 119332 & 7.77 & 5.89 & 0.83 & 5187 & 4.46 & 0.68 &$-$0.03&  3.21 &  0.00 & $-$0.04&  27.9\\
054906 &  97658 & 7.76 & 6.12 & 0.84 & 5175 & 4.58 & 0.61 &$-$0.27&  3.05 & $-$0.25 & $-$0.03&  20.4\\
006379 &   7924 & 7.17 & 6.04 & 0.83 & 5173 & 4.60 & 0.65 &$-$0.15&  2.99 & $-$0.07 & $-$0.03&  25.8\\
007830 &  10261 & 8.92 & 5.84 & 0.91 & 5165 & 4.59 & 0.88 & 0.04&  3.50 &  0.01 & $-$0.03&  27.1\\
015099 &  20165 & 7.83 & 6.09 & 0.86 & 5164 & 4.56 & 0.65 & 0.01&  3.11 &  0.01 & $-$0.03&  25.9\\
073005 & 132142 & 7.77 & 5.88 & 0.79 & 5157 & 4.53 & 0.38 &$-$0.38&  2.98 &  0.00 & $-$0.04&  26.6\\
013891 &  18450 & 8.21 & 5.93 & 0.87 & 5154 & 4.55 & 0.56 &$-$0.06&  3.02 &  0.03 & $-$0.04&  26.4\\
006613 &   8553 & 8.49 & 5.89 & 0.91 & 5129 & 4.61 & 0.59 & 0.00&  2.99 & $-$0.02 & $-$0.03&  24.7\\
112527 & 216520 & 7.53 & 6.03 & 0.87 & 5123 & 4.52 & 0.57 &$-$0.14&  3.12 & $-$0.19 & $-$0.03&  20.8\\
064457 & 114783 & 7.56 & 6.01 & 0.93 & 5121 & 4.47 & 0.69 & 0.13&  3.03 &  0.06 & $-$0.04&  26.5\\
036704 &  59747 & 7.68 & 6.21 & 0.86 & 5120 & 4.60 & 0.85 & 0.03&  3.36 &  0.10 & $-$0.03&  28.0\\
114886 & 219538 & 8.07 & 6.15 & 0.87 & 5110 & 4.56 & 0.65 & 0.01&  2.98 & $-$0.02 & $-$0.03&  24.4\\
090790 & 170657 & 6.81 & 6.21 & 0.86 & 5087 & 4.38 & 0.64 &$-$0.17&  3.45 & $-$0.14 & $-$0.04&  22.2\\
012158 &  16287 & 8.10 & 6.17 & 0.94 & 5081 & 4.54 & 1.00 & 0.14&  3.91 &  0.03 & $-$0.03&  24.0\\
072312 & 130307 & 7.76 & 6.29 & 0.89 & 5078 & 4.55 & 0.75 &$-$0.15&  3.44 & $-$0.18 & $-$0.03&  18.6\\
098505 & 189733 & 7.67 & 6.25 & 0.93 & 5076 & 4.42 & 1.06 & 0.03&  4.14 & $-$0.06 & $-$0.03&  22.5\\
053486 &  94765 & 7.37 & 6.15 & 0.92 & 5076 & 4.59 & 0.94 & 0.06&  3.82 &  0.05 & $-$0.03&  24.0\\
013976 &  18632 & 7.97 & 6.12 & 0.93 & 5075 & 4.59 & 0.98 & 0.19&  3.85 &  0.17 & $-$0.03&  27.5\\
\hline
\end{tabular}
\end{center}
\end{table}

\setcounter{table}{0}
\begin{table}[h]
\scriptsize
\caption{(Continued.)}
\begin{center}
\begin{tabular}{rrrrrrrrrrrrr
}\hline\hline                 
HIP & HD & $V$ & $M_{V}$ & $B-V$ & $T_{\rm eff}$ & $\log g$ & $v_{\rm t}$ & [Fe/H] &  
 $v_{\rm M}$ & [O/H] & $\Delta_{7774}$ & $W_{7774}$ \\
(1) & (2) & (3) & (4) & (5) & (6) & (7) & (8) & (9) & (10) &(11) &(12) &(13) \\
\hline
098828 & 190470 & 7.82 & 6.15 & 0.92 & 5071 & 4.62 & 0.77 & 0.17&  3.04 &  0.14 & $-$0.03&  26.6\\
026505 &  37008 & 7.74 & 6.18 & 0.83 & 5054 & 4.58 & 0.09 &$-$0.41&  2.79 & $-$0.02 &  0.01&  22.8\\
108156 & 208313 & 7.73 & 6.19 & 0.91 & 5051 & 4.59 & 0.72 &$-$0.01&  3.04 &  0.02 & $-$0.03&  22.5\\
108028 & 208038 & 8.18 & 6.28 & 0.94 & 5035 & 4.62 & 0.84 &$-$0.05&  3.57 &  0.01 & $-$0.03&  21.2\\
011083 &  14687 & 8.83 & 6.18 & 0.91 & 5033 & 4.58 & 0.59 & 0.08&  2.94 &  0.04 & $-$0.03&  22.4\\
057939 & 103095 & 6.42 & 6.61 & 0.75 & 5033 & 4.38 &$-$0.10 &$-$1.27&  3.28 & $-$0.93 & $-$0.03&   5.1\\
017420 &  23356 & 7.10 & 6.36 & 0.93 & 5030 & 4.57 & 0.80 &$-$0.08&  3.11 & $-$0.14 & $-$0.03&  17.6\\
088972 & 166620 & 6.38 & 6.15 & 0.88 & 5019 & 4.62 & 0.28 &$-$0.15&  2.73 &  0.05 & $-$0.03&  22.4\\
098792 & 190404 & 7.28 & 6.32 & 0.81 & 5016 & 4.65 &$-$0.18 &$-$0.57&  2.93 & $-$0.16 & $-$0.03&  16.6\\
003206 &   3765 & 7.36 & 6.17 & 0.94 & 5000 & 4.53 & 0.77 & 0.14&  3.02 &  0.15 & $-$0.03&  24.4\\
092919 & 175742 & 8.16 & 6.50 & 0.91 & 4983 & 4.48 & 2.13 &$-$0.10& 11.88 &  0.34 & $-$0.04&  31.6\\
049699 &  87883 & 7.56 & 6.28 & 0.96 & 4980 & 4.62 & 0.56 & 0.11&  2.79 &  0.11 & $-$0.03&  21.7\\
071395 & 128311 & 7.48 & 6.38 & 0.97 & 4967 & 4.67 & 0.88 & 0.16&  4.03 &  0.15 & $-$0.02&  20.9\\
003535 &   4256 & 8.03 & 6.32 & 0.98 & 4954 & 4.47 & 0.75 & 0.26&  3.06 &  0.15 & $-$0.03&  22.7\\
084195 & 155712 & 7.95 & 6.39 & 0.94 & 4947 & 4.43 & 0.57 &$-$0.10&  3.16 & $-$0.07 & $-$0.03&  17.9\\
072146 & 130004 & 7.87 & 6.42 & 0.93 & 4930 & 4.57 & 0.49 &$-$0.24&  2.97 & $-$0.17 & $-$0.02&  13.9\\
033852 &  51866 & 7.98 & 6.43 & 0.99 & 4927 & 4.56 & 0.69 & 0.10&  3.15 &  0.01 & $-$0.02&  17.2\\
084616 & 156985 & 7.93 & 6.59 & 1.02 & 4916 & 4.39 & 0.61 &$-$0.06&  3.31 & $-$0.30 & $-$0.02&  12.0\\
035872 &  57901 & 8.19 & 6.20 & 0.96 & 4908 & 4.58 & 0.53 & 0.17&  2.91 &  0.25 & $-$0.03&  22.4\\
068184 & 122064 & 6.49 & 6.47 & 1.04 & 4908 & 4.49 & 0.67 & 0.23&  3.13 &  0.09 & $-$0.02&  19.0\\
066147 & 117936 & 7.98 & 6.65 & 1.03 & 4872 & 4.28 & 0.95 & 0.01&  3.63 & $-$0.18 & $-$0.02&  13.3\\
071181 & 128165 & 7.24 & 6.60 & 1.00 & 4868 & 4.60 & 0.78 &$-$0.03&  2.97 & $-$0.08 & $-$0.02&  12.8\\
046580 &  82106 & 7.20 & 6.68 & 1.00 & 4861 & 4.59 & 0.88 &$-$0.02&  3.55 &  0.12 & $-$0.02&  16.7\\
000974 &   $\cdots$  & 8.73 & 6.68 & 1.04 & 4852 & 4.67 & 0.57 & 0.02&  3.00 &  0.02 & $-$0.02&  13.8\\
023311 &  32147 & 6.22 & 6.49 & 1.05 & 4815 & 4.49 & 0.66 & 0.29&  2.95 &  0.28 & $-$0.02&  19.4\\
069526 & 124642 & 8.03 & 6.84 & 1.06 & 4798 & 4.51 & 0.90 & 0.10&  4.00 &  0.03 & $-$0.02&  13.2\\
010416 &  13789 & 8.55 & 6.75 & 1.05 & 4782 & 4.64 & 0.86 & 0.09&  3.31 &  0.02 & $-$0.01&  11.6\\
105038 & 202575 & 7.88 & 6.84 & 1.02 & 4777 & 4.66 & 0.75 &$-$0.07&  3.57 &  0.04 & $-$0.01&  11.9\\
032010 &  47752 & 8.08 & 6.86 & 1.02 & 4776 & 4.53 & 0.78 &$-$0.17&  3.16 & $-$0.06 & $-$0.02&  11.2\\
025220 &  35171 & 7.93 & 7.15 & 1.10 & 4757 & 4.43 & 1.06 &$-$0.10&  4.04 & $-$0.20 & $-$0.02&   8.9\\
008275 &  10853 & 8.91 & 7.10 & 1.04 & 4739 & 4.73 & 0.66 &$-$0.10&  3.29 &  0.00 & $-$0.01&   9.8\\
011000 &  14635 & 9.07 & 6.93 & 1.08 & 4732 & 4.71 & 0.70 & 0.19&  3.40 &  0.17 & $-$0.01&  12.1\\
015919 &  21197 & 7.86 & 6.96 & 1.15 & 4717 & 4.22 & 0.89 & 0.14&  3.34 &  0.07 & $-$0.02&  13.3\\
005286 &   6660 & 8.41 & 6.83 & 1.12 & 4716 & 4.45 & 0.72 & 0.20&  3.25 &  0.01 & $-$0.01&  10.7\\
098698 & 190007 & 7.46 & 6.87 & 1.13 & 4677 & 4.50 & 0.79 & 0.22&  3.34 &  0.10 & $-$0.01&  10.5\\
013258 &  17660 & 8.87 & 7.12 & 1.19 & 4643 & 4.32 & 0.67 & 0.23&  3.25 &  0.27 & $-$0.02&  13.5\\
104214 & 201091 & 5.21 & 7.49 & 1.18 & 4523 & 4.57 & 0.32 &$-$0.28&  3.18 & $-$0.15 & $-$0.01&   4.4\\
$\cdots$ &131156B & 6.82 & 7.67 & 1.17 & 4495 & 4.55 & 0.67 &$-$0.25&  3.59 & $-$0.12 & $-$0.01&   4.2\\
\hline
\end{tabular}
\end{center}
(1) Hipparcos Catalog number. (2) Henry-Draper Catalog number. 
(3) Apparent visual magnitude (in mag). (4) Absolute visual magnitude (in mag).
(5) $B-V$ color (in mag). (6) Effective temperature (in K).
(7) Logarithmic surface gravity (cm~s$^{-2}$/dex). (8) Microturbulent velocity 
dispersion (in km~s$^{-1}$). (9) Fe abundance relative to the Sun (in dex).
(10) Macrobroadening velocity (in km~s$^{-1}$). 
(11) Non-LTE oxygen abundance relative to the Sun, $A^{\rm NLTE}$(O)$-8.861$ (in dex), 
where 8.861 is the solar non-LTE oxygen abundance derived in the same manner
(cf. Takeda et al. 2015). (12) Non-LTE correction ($\equiv A^{\rm N} - A^{\rm L}$)
(in dex) for O I 7774.166 (middle line of the triplet).
(13) Equivalent width for O I 7774.166 (in m\AA).\\
In each of the stellar group (47 Hyades stars and 101 field stars), the data are 
arranged in the decreasing order of $T_{\rm eff}$ similarly to Figures 4 and 5,
so that a direct comparison may be possible. (See ``tableE1.dat'' of the online material
for the data arranged in the increasing order of HIP number for each group.)  
\end{table}

\setcounter{table}{1}
\begin{table}[h]
\scriptsize
\caption{Published [O/H] derivations from O~{\sc i} 7771--5 lines for open cluster stars.}
\begin{center}
\begin{tabular}{cccccccl
}\hline\hline                 
Cluster & Ref. & Figure & [O/H]$_{5000}$ & [O/H]$_{6000}$ & 
$T_{\rm eff}$ formula & [Fe/H] & Remark\\  
(1) & (2) & (3) & (4) & (5) & (6) & (7) & (8) \\
\hline
Hyades   & SCH06 & Fig.~3 & $\sim 0.5$ & $\sim 0.2$ & Equation~(A1) & +0.15 & \\
Pleiades & SCH04 & Fig.~1 & $\sim 0.9$ & $\sim 0.2$ & Equation~(A2) & +0.01 & See also Fig.~4 in KS05.\\
M~34     & SCH04 & Fig.~1 & $\sim 0.8$ & $\sim 0.1$ & $\cdots$& +0.07 & Spectroscopic $T_{\rm eff}$. \\
UMa group& KS05  & Fig.~4 & $\sim 0.3$ & $\sim 0.1$ & Equation~(A3) & $-0.08$ & \\
Hyades   & MAD13 & Fig.~4 & $\cdots$   & $\sim 0.2$ & Equation~(A1) & +0.15 & Lowest $T_{\rm eff}$ $\sim 5400$~K, where [O/H]~$\sim 0.3$.\\ 
NGC~752  & MAD13 & Fig.~5 & $\sim 0.7$ & $\sim -0.1$& Equation~(A1) & $-0.06$ & E($B-V$) = 0.035 was adopted.\\
\hline
\end{tabular}
\end{center}
(1) Cluster name. (2) Reference key: SCH06 --- Schuler et al. (2006a), SCH04 --- Schuler et al. (2004),
KS05 --- King \& Schuler (2005), MAD13 --- Maderak et al. (2013). (3) Figure number of the relevant
paper where [O/H] vs. $T_{\rm eff}$ plots for cluster stars are presented. (4) Rough value of [O/H]
at $T_{\rm eff} \sim 5000$~K. (5) Rough value of [O/H] at $T_{\rm eff} \sim 6000$~K. 
(6) $T_{\rm eff}$ vs. $(B-V)_{0}$ formula adopted in the relevant study for evaluation of $T_{\rm eff}$. 
(7) [Fe/H] of the cluster, which we used for evaluation of $T_{\rm eff}^{\rm Casagrande}$ (cf. Figure~14b)
by using Casagrande et al.'s (2010) relation. (8) Specific remark.
\end{table}

\clearpage

\begin{figure}
\epsscale{.65}
\plotone{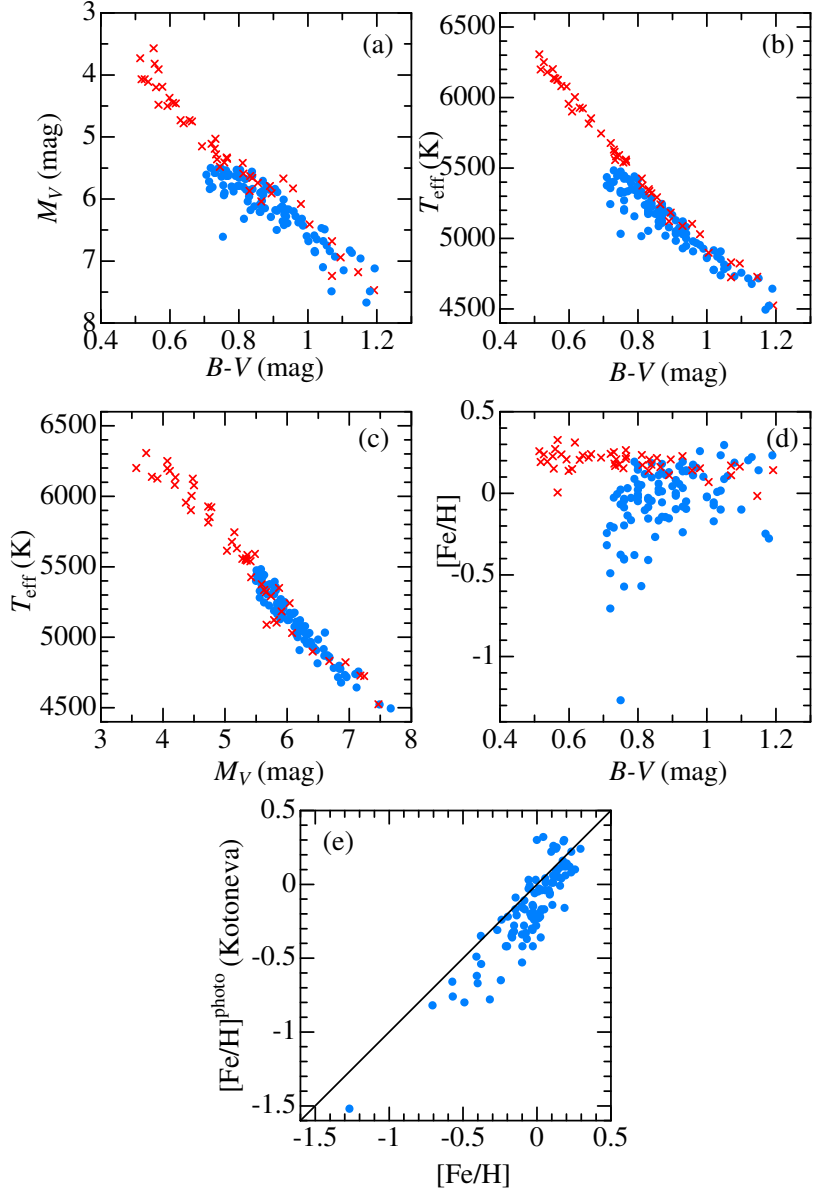}
\caption{
Panels (a)--(d) are the correlation diagram between the photometric data 
(absolute visual magnitude $M_{V}$ and $B-V$ color, mainly derived from 
the Hipparcos catalog) and the spectroscopically determined atmospheric 
parameters (cf. Section 3.1) of the program stars: (a) $M_{V}$ vs. $B-V$,
(b) $T_{\rm eff}$ vs. $B-V$, (c) $T_{\rm eff}$ vs. $M_{V}$. and 
(d) [Fe/H] vs. $B-V$.
Panel (e) shows the interrelationship between the spectroscopic metallicity 
established in this study (abscissa) and the photometric metallicity 
([Fe/H]$^{\rm photo}$) evaluated by Kotoneva et al. (2002) based on 
the position in the color--magnitude diagram (ordinate).
Our program stars of 47 Hyades stars and 101 field stars are separately 
represented by red crosses and blue circles, respectively. 
}
\end{figure}

\clearpage

\begin{figure}
\epsscale{.65}
\plotone{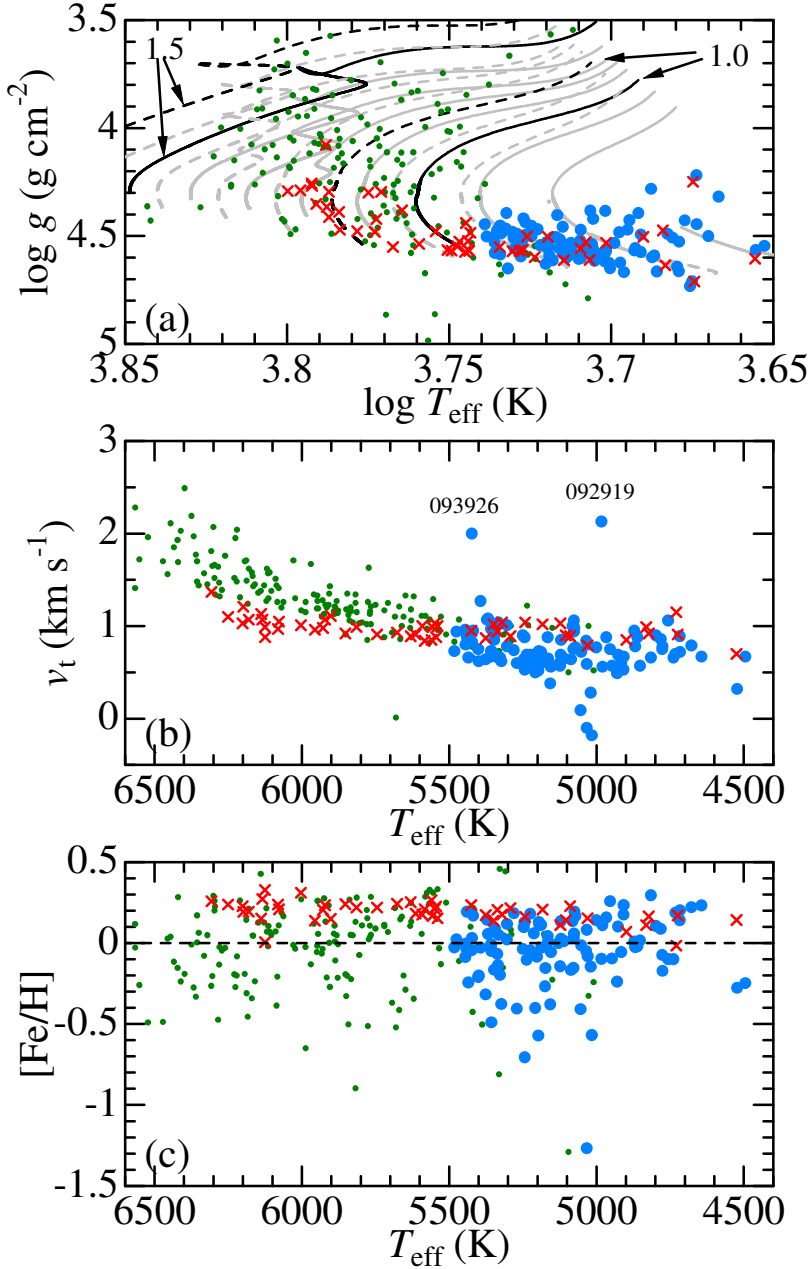}
\caption{
Spectroscopically determined $\log g$, $v_{\rm t}$, 
and [Fe/H] are plotted against $T_{\rm eff}$ in panels (a), (b), and (c),
respectively. In panel (a) are also depicted the theoretical $\log g$ vs.
$\log T_{\rm eff}$ relations corresponding to eight different masses (0.7, 0.8, 0.9, 1.0.
1.1, 1.2, 1.3, 1.4, and 1.5 $M_{\odot}$) for different metallicities
($z = 0.01$ and $z = 0.02$ in dashed and solid lines, respectively), 
which were taken from PARSEC stellar evolutionary tracks
(Bressan et al. 2012, 2013). Apart from the program stars of this study
(47 Hyades and 101 field stars shown by blue circles and red crosses 
respectively), 160 mid-F through early K dwarfs/subgiants investigated 
by Takeda et al. (2005) are also plotted in green dots for comparison.
}
\end{figure}

\clearpage

\begin{figure}
\epsscale{.45}
\plotone{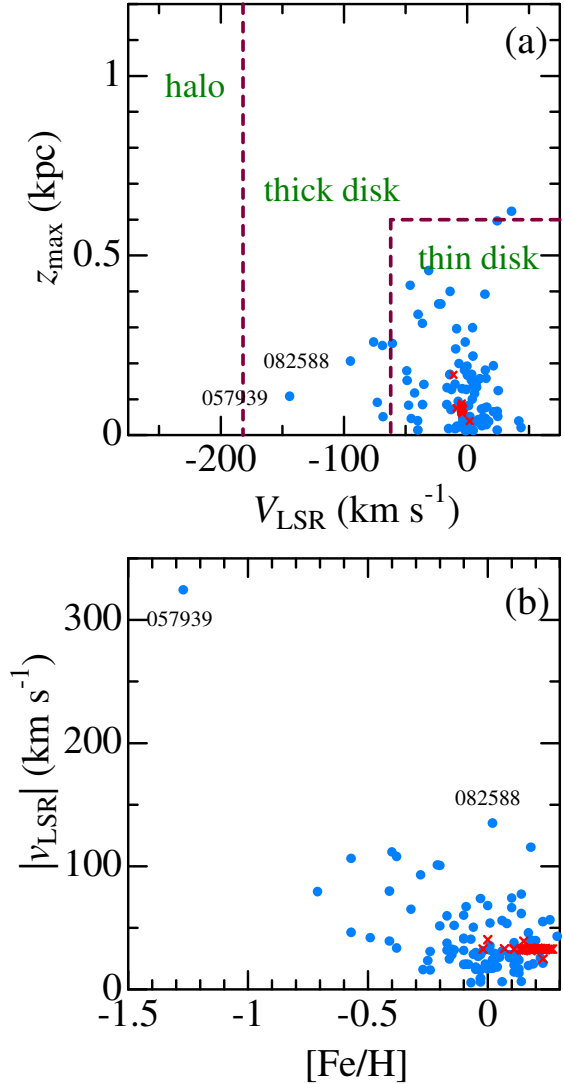}
\caption{
(a) Correlation diagram between the maximum separation from the 
galactic plane ($z_{\rm max}$) and the rotation velocity component 
relative to the Local Standard of Rest ($V_{\rm LSR}$), which may be 
used for classifying the stellar population (the boundaries are 
indicated by the dashed lines; cf. Ibukiyama \& Arimoto 2002).
(b) Space velocity relative to the Local Standard of Rest  
[$|v_{\rm LSR}| \equiv (U_{\rm LSR}^{2} + V_{\rm LSR}^{2} + W_{\rm LSR}^{2})^{1/2}$] 
plotted against [Fe/H]. Same meanings of the symbols as in Figure~1. 
}
\end{figure}

\clearpage

\begin{figure}
\epsscale{.65}
\plotone{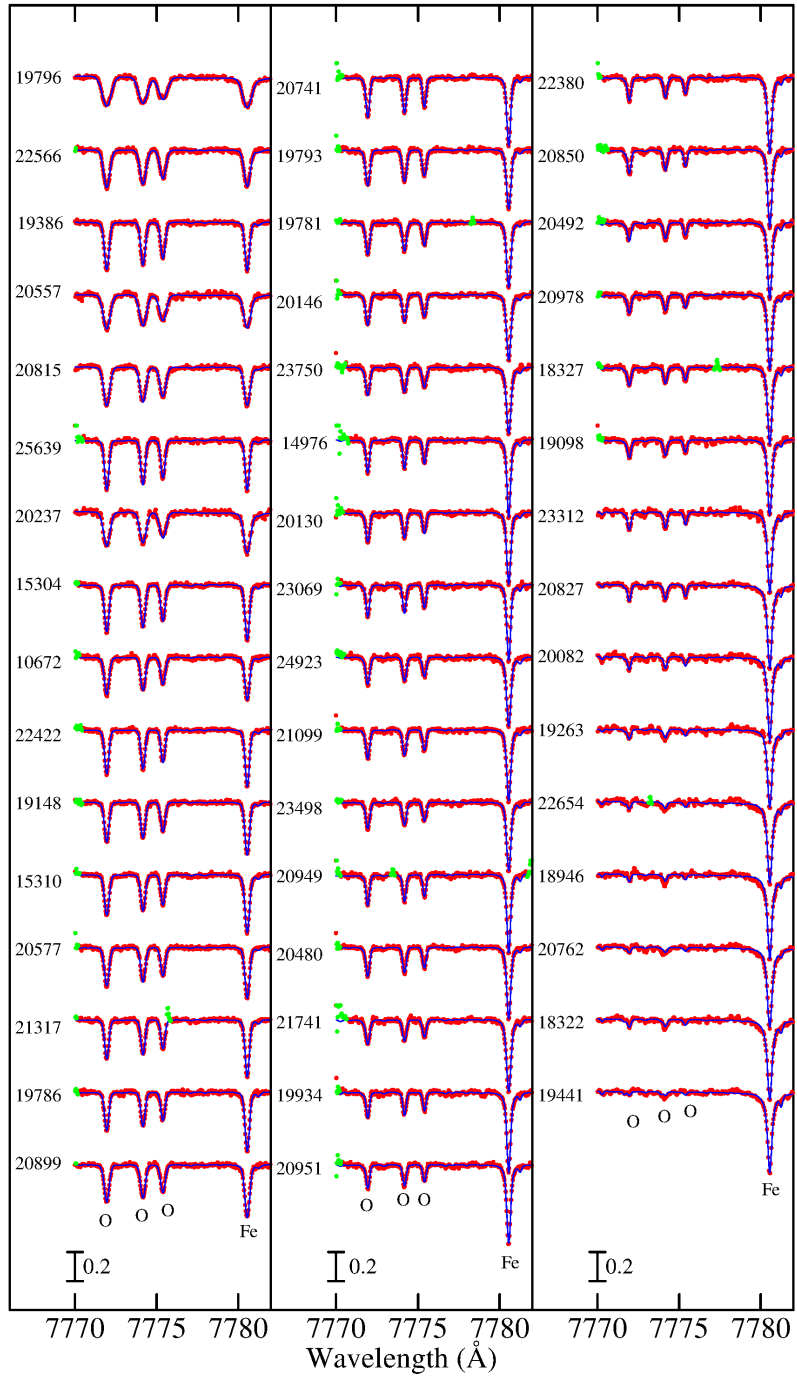}
\caption{
Synthetic spectrum fitting in the 7770--7782~\AA\ region 
comprising the O~{\sc i}~7771--5 and Fe~{\sc i} 7780 lines
for 47 Hyades stars. 
The best-fit theoretical spectra are shown by dark blue solid lines. 
The observed data used in the fitting are plotted by red symbols, 
while those rejected in the fitting (e.g., due to spectrum defect) 
are highlighted in green.
In each panel (from left to right), the spectra are arranged in the 
descending order of $T_{\rm eff}$ as in Table~1, and vertical offsets 
of 0.5 are applied to each spectrum (indicated by the HIP number) 
relative to the adjacent one. The wavelength scale is adjusted to 
the laboratory frame by correcting the stellar radial velocity.
}
\end{figure}

\clearpage

\begin{figure}
\epsscale{.65}
\plotone{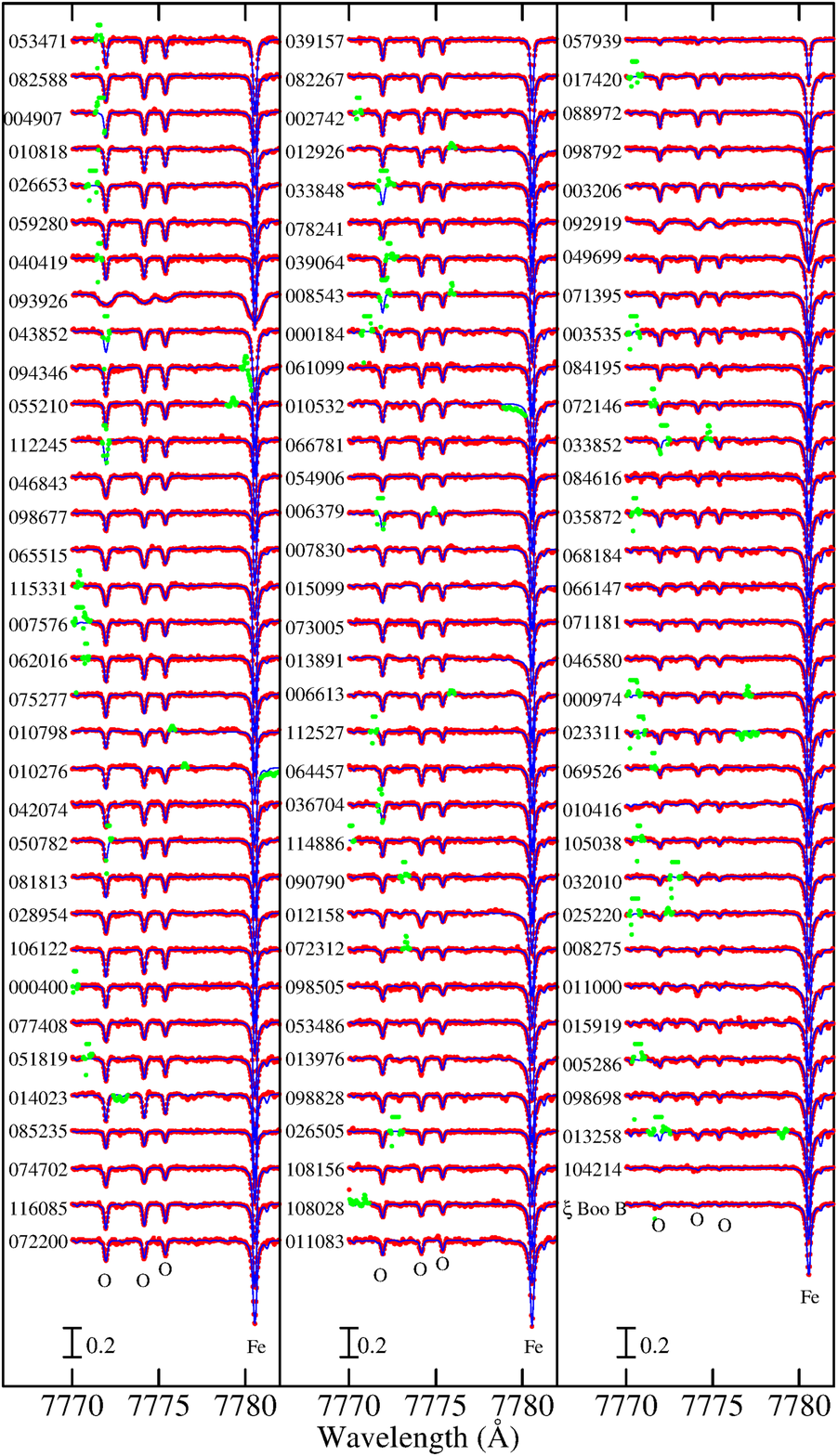}
\caption{
Synthetic spectrum fitting in the 7770--7782~\AA\ region 
for 101 field stars. A vertical offset of 0.25 is applied to
each spectrum relative to the adjacent one. Otherwise, 
the same as in Figure~4.
}
\end{figure}

\clearpage

\begin{figure}
\epsscale{.60}
\plotone{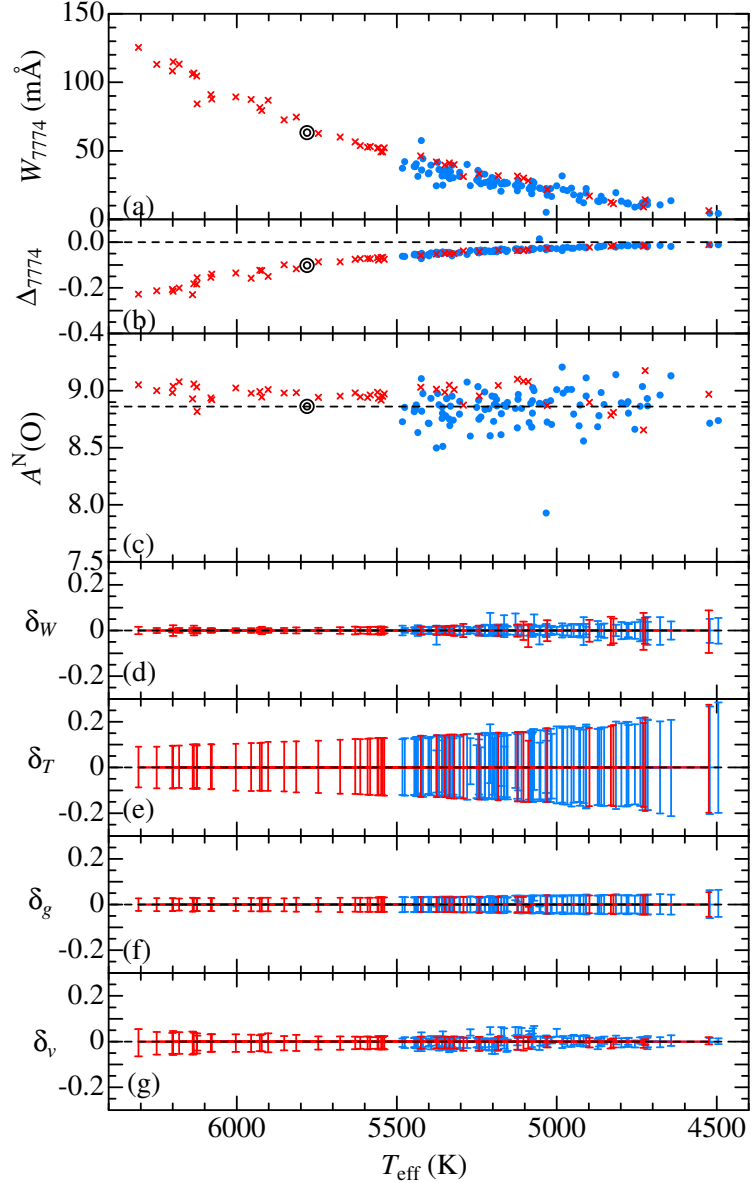}
\caption{
Oxygen abundance and related quantities plotted against $T_{\rm eff}$. 
(a) $W_{7774}$ (equivalent width of O~{\sc i} 7774.166), 
(b) $\Delta_{7774}$ (non-LTE correction for O~{\sc i} 7774.166),
(c) $A^{\rm N}$(O) (non-LTE oxygen abundance derived from spectrum fitting).
(d) $\delta_{W+}$ and $\delta_{W-}$ (abundance change corresponding to
perturbation of $+\delta W$ and $-\delta W$, where $\delta W$ is the uncertainty
of equivalent width evaluated according to Cayrel 1988).
(e) $\delta_{T+}$ and $\delta_{T-}$ (abundance variations 
in response to $T_{\rm eff}$ changes of +100~K and $-100$~K), 
(f) $\delta_{g+}$ and $\delta_{g-}$ (abundance variations 
in response to $\log g$ changes by $+0.1$~dex and $-0.1$~dex), 
and (g) $\delta_{v+}$ and $\delta_{v-}$ (abundance 
variations in response to perturbing the $v_{\rm t}$ value
by +0.5~km~s$^{-1}$ and $-$0.5~km~s$^{-1}$).
Note that the signs of these $\delta$ values are 
$\delta_{W+} > 0$,  $\delta_{W-} < 0$, 
$\delta_{T+} < 0$,  $\delta_{T-} > 0$, 
$\delta_{g+} > 0$,  $\delta_{g-} < 0$, 
$\delta_{v+} < 0$ and  $\delta_{v-} > 0$.
The non-LTE solar O abundance of 8.861 derived in the similar manner (cf. 
Takeda et al. 2015) is indicated by the horizontal dashed line in panel (c). 
The large double circles in panels (a)--(c) denote the solar values (cf. footnote~7).
Otherwise, the same meanings of the symbols (and their colors) as in Figure~1.
}
\end{figure}

\clearpage

\begin{figure}
\epsscale{.70}
\plotone{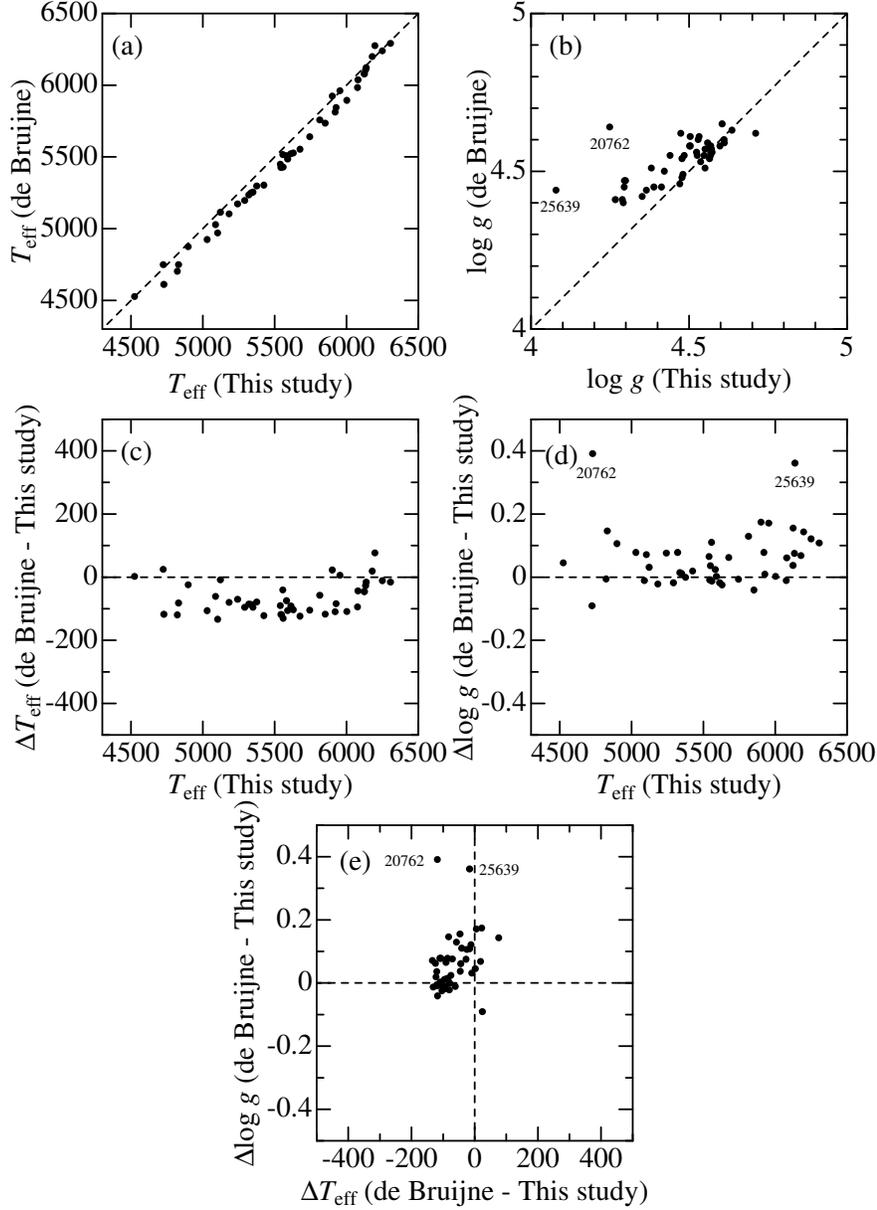}
\caption{
Comparison of the $T_{\rm eff}$ and $\log g$ values determined for the 
Hyades stars in this study (based on Fe~{\sc i} and Fe~{\sc ii} lines)
with those of de Bruijne et al. (2001) 
(all of our 47 Hyades stars are included their sample).
(a) $T_{\rm eff}$(theirs) vs. $T_{\rm eff}$(ours), 
(b) $\log g$(theirs) vs. $\log g$(ours),
(c) $\Delta T_{\rm eff}$(theirs$-$ours) vs. $T_{\rm eff}$(ours),
(d) $\Delta \log g$(theirs$-$ours) vs. $\log g$(ours), and
(e) $\Delta \log g$(theirs$-$ours) vs. $\Delta T_{\rm eff}$(theirs$-$ours).
}
\end{figure}

\clearpage

\begin{figure}
\epsscale{.70}
\plotone{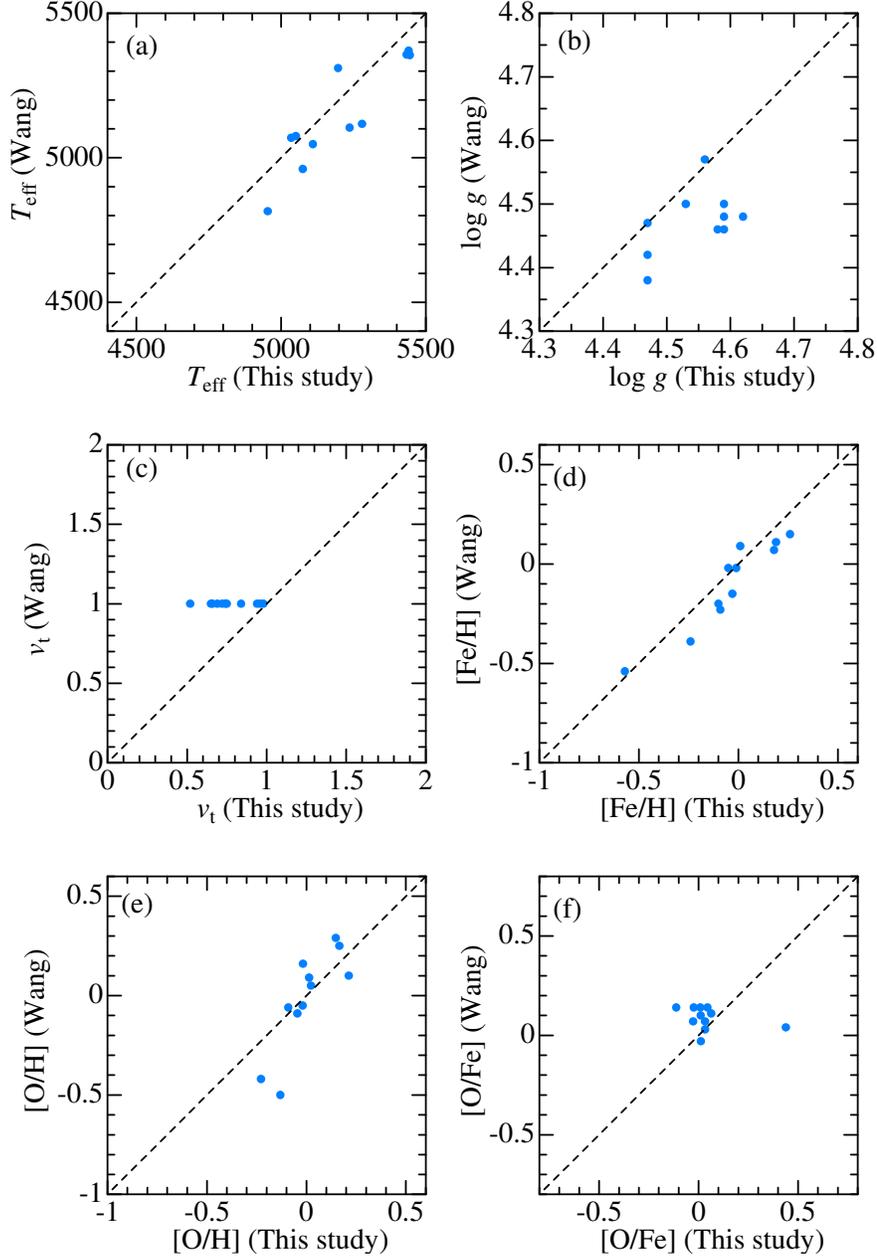}
\caption
{
Comparison of the atmospheric parameters and the oxygen abundances derived
in this study with those of Wang et al. (2009) for 11 field stars in common. 
(a) $T_{\rm eff}$, (b) $\log g$, (c) $v_{\rm t}$, (d) [Fe/H]. (e) [O/H], 
and (f) [O/Fe]. (Note that they assumed $v_{\rm t} =1$~km~s$^{-1}$.) 
}
\end{figure}

\clearpage

\begin{figure}
\epsscale{.70}
\plotone{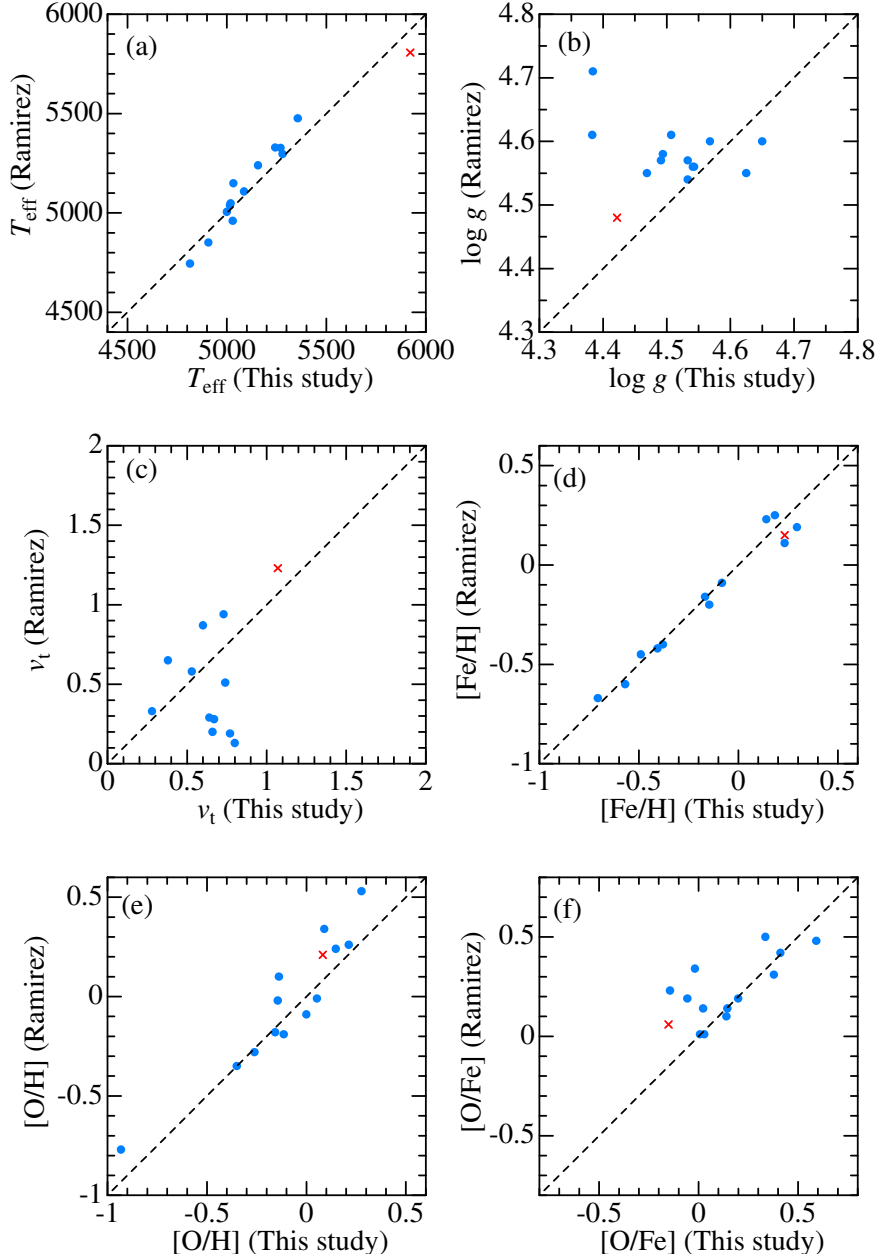}
\caption
{
Comparison of the atmospheric parameters and the oxygen abundances derived
in this study with those of Ram\'{\i}rez et al. (2013) for 13 field stars and 
1 Hyades star in common. Otherwise, the same as in Figure~8. 
}
\end{figure}

\clearpage

\begin{figure}
\epsscale{.70}
\plotone{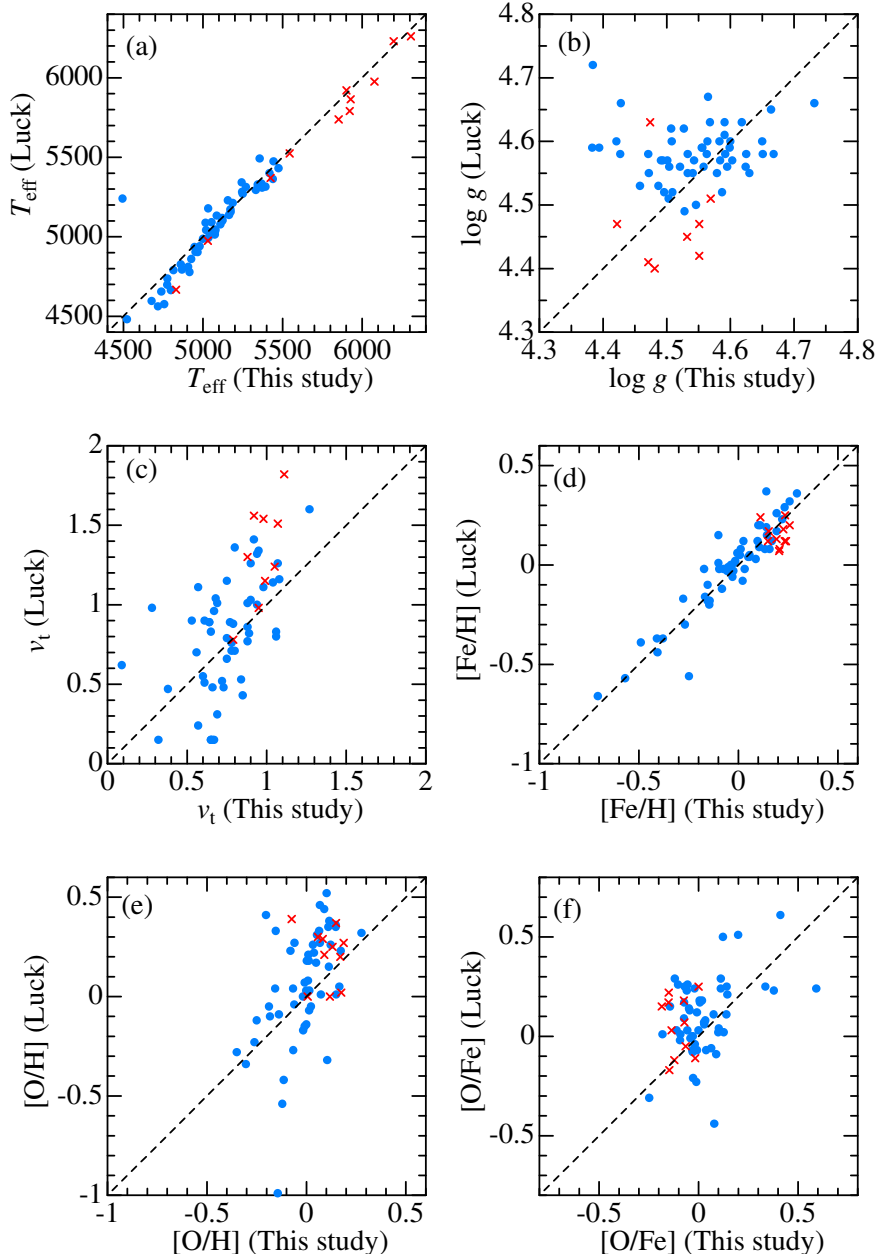}
\caption
{
Comparison of the atmospheric parameters and the oxygen abundances derived
in this study with those of Luck (2017) for 55 field stars and 11 Hyades 
stars in common. Otherwise, the same as in Figure~8.
}
\end{figure}

\begin{figure}
\epsscale{.70}
\plotone{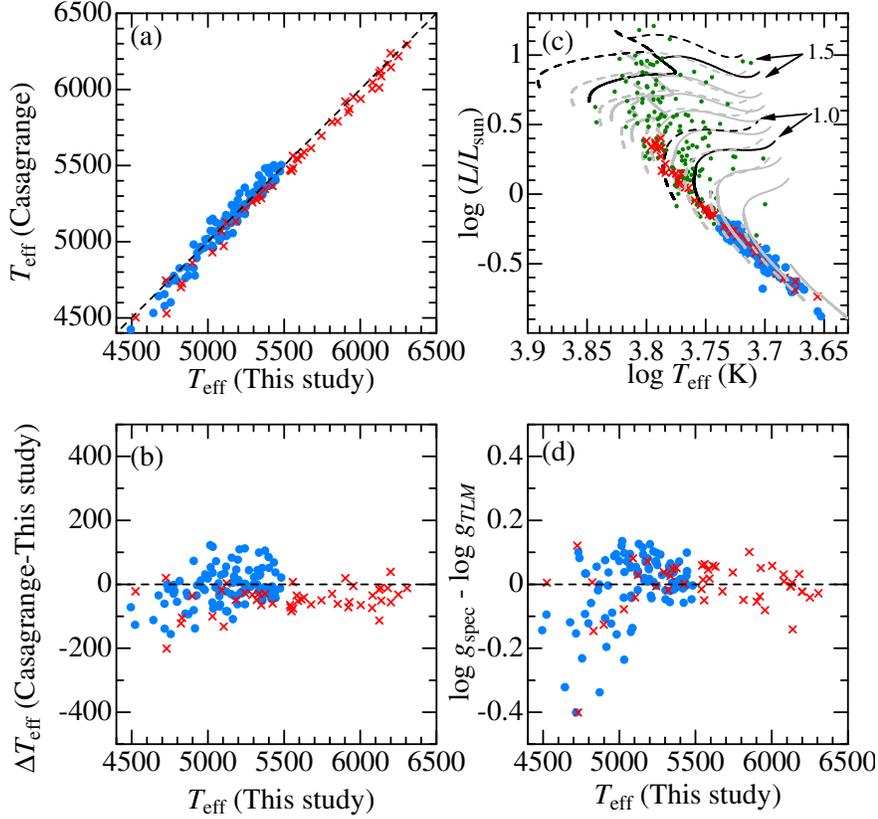}
\caption
{
Panels (a) and (b) illustrate the comparison of the photometric $T_{\rm eff}$ 
derived from $B-V$ and [Fe/H] by using Casagrande et al.'s (2010) formula
with the spectroscopic $T_{\rm eff}$ adopted in this study:
(a) $T_{\rm eff}$(Casagrande) vs. $T_{\rm eff}$(This study) and 
(b) $T_{\rm eff}$(Casagrande)$-T_{\rm eff}$(This study) vs. $T_{\rm eff}$(This study).
In panel (c) is shown the $L$(bolometric luminosity) vs. $T_{\rm eff}$ relation
for the program stars, where the theoretical PARSEC tracks are also depicted 
similarly to Figure~2a. The differences between $\log g_{\rm spec}$ (spectroscopic 
surface gravity adopted in this study) and $\log g_{TLM}$ (theoretical surface gravity
derived from $T_{\rm eff}$, $L$, and $M$) are plotted against $T_{\rm eff}$ in panel (d).   
See the caption of Figure~2 for the meanings of the symbols and lines.
}
\end{figure}

\clearpage

\begin{figure}
\epsscale{.70}
\plotone{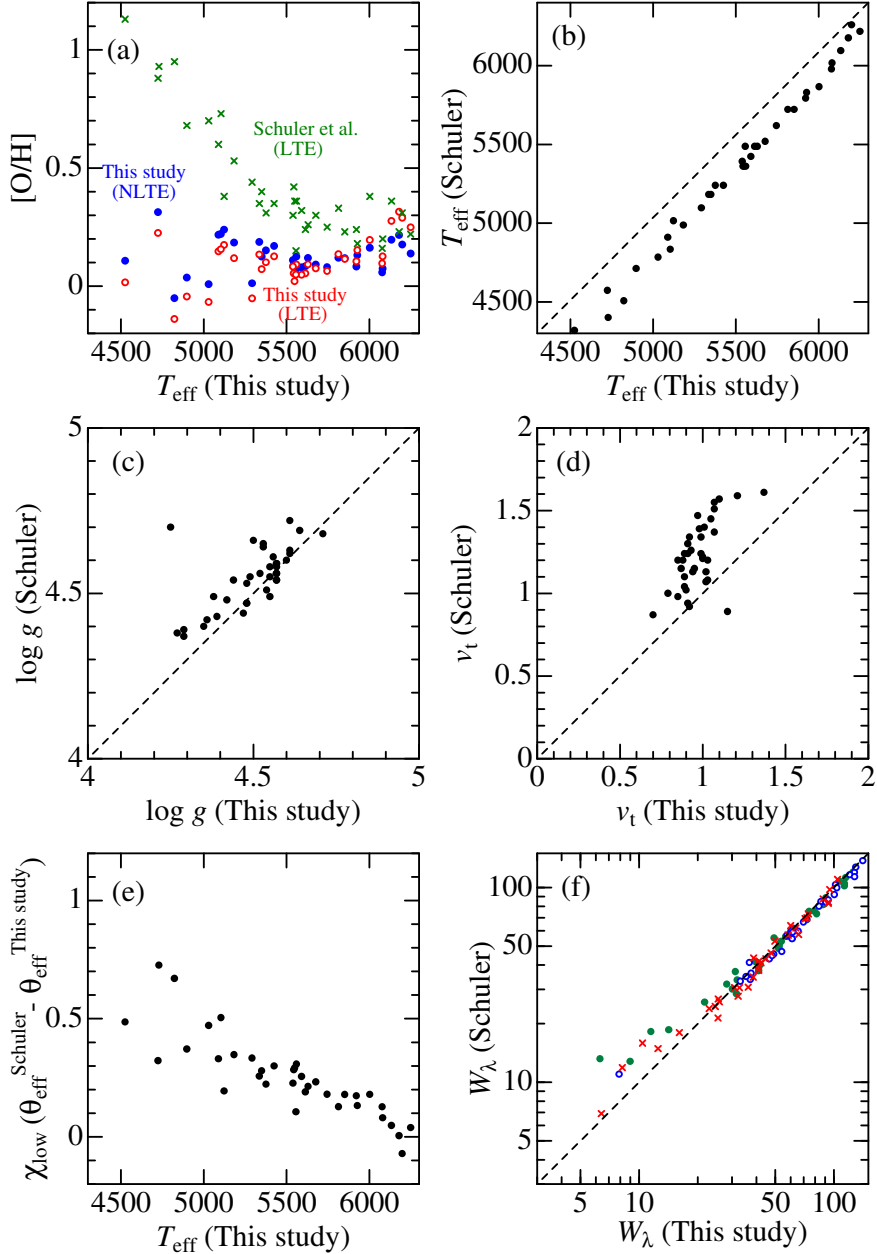}
\caption{
Comparison of the our oxygen abundances and atmospheric parameters derived for 
Hyades stars with those of Schuler et al. (2006a) (37 stars are in common). 
(a) [O/H] values plotted against $T_{\rm eff}$(ours), where 
[O/H](theirs, LTE), [O/H](ours, NLTE), ad [O/H](ours, LTE)
are denoted by green crosses, blue filled circles, and 
red open circles, respectively.
(b) $T_{\rm eff}$(theirs) vs. $T_{\rm eff}$(ours). 
(c) $\log g$(theirs) vs. $\log g$(ours).
(d) $v_{\rm t}$(theirs) vs. $v_{\rm t}$(ours).
(e) Differences of 
$\chi_{\rm low}\theta_{\rm eff}$(theirs)$-$$\chi_{\rm low}\theta_{\rm eff}$(ours)
(which defines the shift in the abscissa of curve of growth) plotted against 
$T_{\rm eff}$(ours), where $\chi_{\rm low}$ = 9.146~eV (lower excitation potential 
of the O~{\sc i} 7771--5 triplet), and $\theta_{\rm eff} \equiv 5040/T_{\rm eff}$
($T_{\rm eff}$ in K).
(f) $W$(theirs) vs. $W$(ours) diagram, where blue open circles, green filed circles, 
and red crosses correspond to O~{\sc i} 7771.944, 7774.166, and 7775.388 lines, 
respectively. 
}
\end{figure}

\clearpage

\begin{figure}
\epsscale{.50}
\plotone{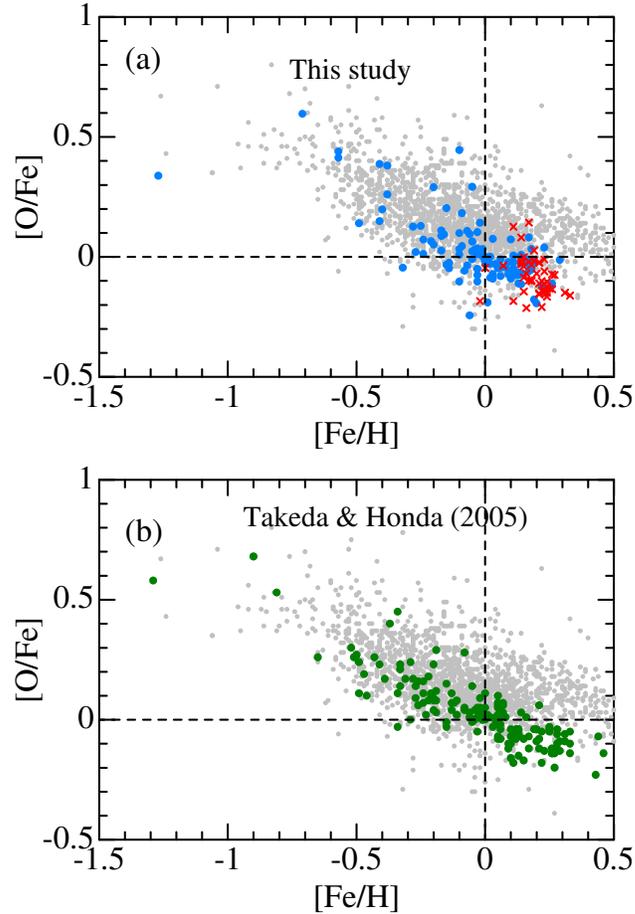}
\caption
{
(a) The [O/Fe] ratios derived in this study for each of the program stars (47 Hyades 
stars and 101 field stars) are plotted against [Fe/H], where the meanings of the 
symbols are the same as in Figure~1. 
(b) Takeda \& Honda's (2005) [O/Fe] vs. [Fe/H] relation derived for the 160 mid-F 
through early K stars based on the O~{\sc i} 7771--5 lines. 
In both panels (a) and (b) are also plotted the similar correlations 
taken from Hawkins et al. (2016) by gray dots for comparison,
which were derived from the high-resolution infrared spectra 
for a large APOGEE+Kepler stellar sample (APOKASC).
}
\end{figure}

\clearpage

\begin{figure}
\epsscale{.40}
\plotone{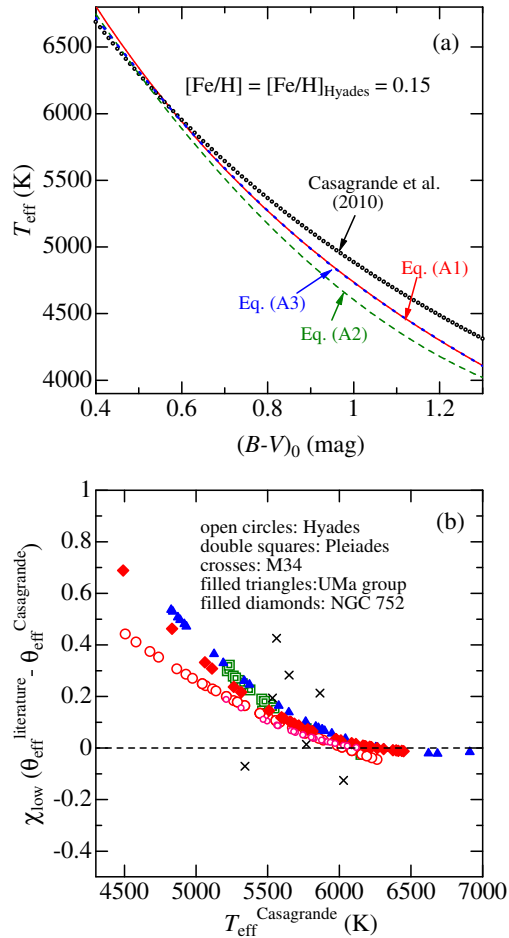}
\caption
{
(a) Comparison of three $T_{\rm eff}$ vs. $B-V$ relations (lines) used in previous
studies on the [O/H] trends of open cluster stars (cf. Table~2)
with that of Casagrande et al. (2010) (black filled circles), where [Fe/H] is set
at +0.15 (= [Fe/H]$_{\rm Hyades}$) for all cases. Red solid line --- Equation~(A1),
green dashed line --- Equation~(A2), and blue dotted line --- Equation~(A3).
(b) Differences of 
$\chi_{\rm low}\theta_{\rm eff}$(literature)$-$$\chi_{\rm low}\theta_{\rm eff}$(Casagrande)
(see the caption of Figure~12e for the meanings of $\theta_{\rm eff}$ and $\chi_{\rm low}$) 
are plotted against $T_{\rm eff}$(Casagrande), where  $T_{\rm eff}$(literature) is
the actual value used by the relevant study and $T_{\rm eff}$(Casagrande) was
calculated according to Casagrande et al. (2010) along with each star's 
$B-V$ and cluster [Fe/H]. Large open circles --- Hyades (SCH06), small open circles
--- Hyades (MAD13), double squares --- Pleiades (SCH04), crosses --- M~34 (SCH04),
filled triangles --- UMa moving group (KS05), and filled diamonds --- NGC~752 (MAD13).
(See the caption of Table~2 for the key to the reference code.) Note that the colors of 
the symbols are so chosen as to match those of the lines in panel (a) corresponding to
the $T_{\rm eff}$ vs. $B-V$ formula adopted in each paper.  
}
\end{figure}

\clearpage






\clearpage





\end{document}